\documentclass[longauth]{aa}  
\usepackage{graphicx}
\usepackage{longtable}
\usepackage{tabularx}
\usepackage{txfonts}
\usepackage{hyperref}
\usepackage{amsmath}
\usepackage{scalerel}
\usepackage{natbib}
\usepackage{tablefootnote}

\usepackage{xcolor}
\definecolor{dark-red}{rgb}{0.9,0.0,0.0}
\definecolor{dark-blue}{rgb}{0.15,0.15,0.9}
\definecolor{dark-green}{rgb}{0.15,0.8,0.15}
\definecolor{medium-blue}{rgb}{0,0,0.9}
\hypersetup{
    colorlinks, linkcolor=red,
    citecolor={dark-blue} , urlcolor={medium-blue}
}

\newcommand{\feh}{$[$Fe/H$]$\,}

\newcommand{\teff}{\ensuremath{T_{\rm eff}}}

\newcommand{\msun}{M$_\odot$\,}
\newcommand{\mstar}{M$_\star$\,}
\newcommand{\rstar}{R$_\star$\,}
\newcommand{\lstar}{$L_\star$\,}
\newcommand{\rsun}{R$_\odot$\,}
\newcommand{\lsun}{L$_\odot$\,}

\newcommand{\ms}{m\,s$^{-1}$\,}
\newcommand{\mjup}{M$_{\rm J}$\,}
\newcommand{\rjup}{R$_{\rm J}$\,}
\newcommand{\vsini}{$\varv$\,sin\,$i$\,}
\newcommand{\mearth}{M$_\oplus$\,} 
\newcommand{\zaspe}{\texttt{ZASPE}\,}
\newcommand{\SPECIES}{\texttt{SPECIES}\,}
\newcommand{\ceres}{\texttt{CERES}\,}
\newcommand{\parsec}{\texttt{PARSEC}\,}
\newcommand{\tess}{TESS }
\newcommand{\celerite}{\texttt{celerite}}
\newcommand{\mesa}{\texttt{MESA}\,}
\newcommand{\juliet}{\texttt{juliet}\,}
\newcommand{\radvel}{\texttt{radvel}\,}
\newcommand{\batman}{\texttt{batman}\,}
\newcommand{\dynesty}{\texttt{dynesty}\,}
\newcommand{\gaia}{\texttt{GAIA}\,}
\newcommand{\perone}{69.0480$\substack{+0.0004 \\ -0.0005}$\,}
\newcommand{\massone}{12.74$\substack{+1.01 \\ -1.01}$\,}
\newcommand{\radone}{1.026$\substack{+0.065 \\ -0.067}$\,}
\newcommand{\eccone}{0.018$\substack{+0.004 \\ -0.004}$\,}
\newcommand{\pertwo}{64.5949$\substack{+0.0003 \\ -0.0003}$\,}
\newcommand{\masstwo}{2.340$\substack{+0.197 \\ -0.195}$\,}
\newcommand{\radtwo}{1.030$\substack{+0.050 \\ -0.050}$\,}
\newcommand{\ecctwo}{0.021$\substack{+0.024 \\ -0.015}$\,}

\begin{document} 
\title{A long-period transiting substellar companion in the super-Jupiters to brown dwarfs mass regime and a prototypical warm-Jupiter detected by \tess.\thanks{ Based on observations collected at La Silla - Paranal Observatory under programs IDs 105.20GX.001, 106.212H.001, 106.21ER.001 and 108.22A8.001 and through the Chilean Telescope Time under programs IDs CN2020B-21, CN2021A-14, CN2021B-23, CN2022A-33 and CN2022B-33}}
\titlerunning{Two Warm Jupiters detected by TESS}
\authorrunning{M. I. Jones et al.}
\author{
        Mat\'ias I. Jones \inst{1}
        \and Yared Reinarz \inst{2,4,15}
        \and Rafael Brahm \inst{3,4,5}
        \and Marcelo Tala Pinto \inst{3,4}
        \and Jan Eberhardt \inst{15}
        \and Felipe Rojas \inst{12}
        \and Amaury H. M. J. Triaud \inst{14}
        \and Arvind F.\ Gupta \inst{7,8}
        \and Carl Ziegler \inst{9}
        \and Melissa J.\ Hobson\inst{4,15}
        \and Andr\'es Jord\'an \inst{3,4,5}
        \and Thomas Henning \inst{15}
        \and Trifon Trifonov \inst{15,16}
        \and Martin Schlecker \inst{17}
        \and N\'estor Espinoza \inst{18}
        \and Pascal Torres-Miranda \inst{12}
        \and Paula Sarkis \inst{15}
        \and Sol\`ene Ulmer-Moll \inst{23,24}
        \and Monika Lendl \inst{23}
        \and Murat Uzundag \inst{6}        
        \and Maximiliano Moyano \inst{2}
        \and Katharine Hesse \inst{19}
        \and Douglas A. Caldwell \inst{32,33}
        \and Avi Shporer \inst{20}
        \and Michael B. Lund \inst{39}
        \and Jon M. Jenkins \inst{32}
        \and Sara Seager \inst{20,35,36}
        \and Joshua N. Winn \inst{37}
        \and George R. Ricker \inst{20}
        \and Christopher J. Burke \inst{20}
        \and Pedro Figueira \inst{23}
        \and Angelica Psaridi \inst{23}
        \and Khaled Al Moulla \inst{23}
        \and Dany Mounzer \inst{23}
        \and Matthew R. Standing \inst{21}
        \and David V. Martin \inst{22}
        \and Georgina Dransfield \inst{14}
        \and Thomas Baycroft \inst{14}
        \and Diana Dragomir \inst{27}
        \and Gavin Boyle \inst{30,38}
        \and Vincent Suc \inst{3,4,38}
        \and Andrew W. Mann \inst{11}
        \and Mathilde Timmermans \inst{28}
        \and Elsa Ducrot \inst{29}
        \and Matthew J. Hooton \inst{30}
        \and Sebasti\'an Zu\~niga-Fern\'andez \inst{28}
        \and Daniel Sebastian \inst{14}
        \and Michael Gillon \inst{28}
        \and Didier Queloz \inst{30,31}
        \and Joe Carson \inst{13}
        \and Jack J. Lissauer \inst{32}
}

\institute{European Southern Observatory, Alonso de C\'ordova 3107, Vitacura, Casilla, 19001, Santiago, Chile
\and Instituto de Astronom\'ia, Universidad Cat\'olica del Norte, Angamos 0610, 1270709, Antofagasta, Chile
\and Facultad de Ingenier\'ia y Ciencias, Universidad Adolfo Ib\'{a}\~{n}ez, Av. Diagonal las Torres 2640, Pe\~{n}alol\'{e}n, Santiago, Chile
\and Millennium Institute for Astrophysics, Chile 
\and Data Observatory Foundation, Chile
\and Institute of Astronomy, KU Leuven, Celestijnenlaan 200D, B-3001 Leuven, Belgium
\and Department of Astronomy \& Astrophysics, 525 Davey Laboratory, The Pennsylvania State University, University Park, PA, 16802, USA 
\and Center for Exoplanets and Habitable Worlds, 525 Davey Laboratory, The Pennsylvania State University, University Park, PA, 16802, USA
\and Department of Physics, Engineering and Astronomy, Stephen F. Austin State University, 1936 North St, Nacogdoches, TX 75962, USA
\and Cerro Tololo Inter-American Observatory, Casilla 603, La Serena, Chile
\and Department of Physics and Astronomy, The University of North Carolina at Chapel Hill, Chapel Hill, NC 27599-3255, USA
\and Instituto de Astrof\'isica, Facultad de F\'isica, Pontificia Universidad Cat\'olica de Chile, Santiago, Chile
\and Department of Physics \& Astronomy, College of Charleston, 66 George Street Charleston, SC 29424, USA
\and School of Physics \& Astronomy, University of Birmingham, Edgbaston, Birmingham B15 2TT, UK
\and Max-Planck-Institut f\"{u}r Astronomie, K\"{o}nigstuhl  17, 69117 Heidelberg, Germany
\and Department of Astronomy, Sofia University ``St Kliment Ohridski'', 5 James Bourchier Blvd, BG-1164 Sofia, Bulgaria
\and Steward Observatory and Department of Astronomy, The University of Arizona, Tucson, AZ 85721, USA
\and Space Telescope Science Institute, 3700 San Martin Drive, Baltimore, MD 21218, USA
\and Wesleyan University, Middletown, CT 06459, USA
\and Department of Physics and Kavli Institute for Astrophysics and Space Research, Massachusetts Institute of Technology, Cambridge, MA 02139, USA
\and European Space Agency (ESA), European Space Astronomy Centre (ESAC), Camino Bajo del Castillo s/n, 28692 Villanueva de la Ca\~nada, Madrid, Spain
\and Department of Astronomy, The Ohio State University, Columbus, OH, USA
\and Observatoire Astronomique de l'Université de Genève, Chemin Pegasi 51, 1290, Versoix, Switzerland 
\and Physikalisches Institut, University of Bern, Gesellschaftsstrasse 6, 3012, Bern, Switzerland
\and Lowell Observatory, Flagstaff, AZ, USA 
\and Department of Astronomy and Planetary Science, Northern Arizona University, Flagstaff, AZ, USA
\and Department of Physics and Astronomy, University of New Mexico, Albuquerque, NM, USA
\and Astrobiology Research Unit, University of Liège, Allée du 6 août 19, B-4000 Liège (Sart-Tilman), Belgium
\and AIM, CEA, CNRS, Université Paris-Saclay, Université de Paris, F-91191 Gif-sur-Yvette, France
\and Cavendish Laboratory, JJ Thomson Avenue, Cambridge CB3 0HE, UK
\and Department of Physics, ETH Zurich, Wolfgang-Pauli-Strasse 2, CH-8093 Zurich, Switzerland
\and NASA Ames Research Center, Moffett Field, CA 94035, USA
\and SETI Institute, Mountain View, CA 94043 USA
\and Center for Astrophysics | Harvard \& Smithsonian, 60 Garden Street, Cambridge, MA 02138, USA
\and Department of Earth, Atmospheric and Planetary Sciences, Massachusetts Institute of Technology, Cambridge, MA 02139, USA
\and Department of Aeronautics and Astronautics, Massachusetts Institute of Technology, 77 Massachusetts Avenue, Cambridge, MA 02139, USA
\and Department of Astrophysical Sciences, Princeton University, Princeton, NJ 08544, USA
\and El Sauce Observatory -- Obstech, Chile
\and NASA Exoplanet Science Institute, Caltech/IPAC, Mail Code 100-22,1200 E. California Blvd., Pasadena, CA 91125, USA
}

\date{}

\abstract{We report on the confirmation and follow-up characterization of two long-period transiting substellar companions on low-eccentricity orbits around TIC\,4672985 and TOI-2529, whose transit events were detected by the \tess space mission.
Ground-based photometric and spectroscopic follow-up from different facilities, 
confirmed the substellar nature of TIC\,4672985\,$b$, a massive gas giant, in the transition between the super-Jupiters and brown-dwarfs mass regime. 
From the joint analysis we derived the following orbital parameters: $P$ = \perone d, M$_p$ = \massone \mjup, R$_p$ = \radone \rjup and $e$ = \eccone. In addition, the RV time series revealed a significant trend at the $\sim$ 350 \ms\,yr$^{-1}$ level, which is indicative of the presence of a massive outer companion in the system. 
TIC\,4672985\,$b$ is a unique example of a transiting substellar companion with a mass above the deuterium-burning limit, located beyond 0.1 AU and in a nearly circular orbit. These planetary properties are difficult to reproduce from canonical planet formation and evolution models.
For TOI-2529\,$b$, we obtained the following orbital parameters: $P$ = \pertwo d, M$_p$ = \masstwo \mjup, R$_p$ = \radtwo \rjup and $e$ = \ecctwo, making this object a new example of a growing population of transiting warm giant planets.
 
}

\maketitle

\section{Introduction} \label{sec:intro}
Giant planets (M$_p$ $\gtrsim$ 0.3 \mjup) with orbital period between $\sim$ 10-200 $d$, also known as warm Jupiters (WJs), are excellent targets for measuring their bulk composition (e.g. \citealt{fortney2007}), characterizing their atmospheric abundances, and studying their formation and evolution mechanisms (e.g., \citealt{knierim2022}). In particular, they are less affected by the strong stellar irradiation and tidal interactions with the host star, than their innermost siblings, the so-called hot Jupiters.
It is already well established that the strong insolation (f$_p \gtrsim$ 2$\times 10^8$ [erg\,s$^{-1}$\,cm$^{-2}$]; \citealt{demory_seager2011}) received by hot Jupiters is associated with abnormally large radii, known as radius inflation (e.g, \citealt{baraffe2010}; \citealt{laughlin2011}), which in turn are due to delayed cooling and/or energy deposition in the planetary interior. The latter process might even lead to a planet reinflation after the host star evolves off the main sequence \citep{grunblatt2017,komacek2020}. 
However, although different mechanisms have been proposed to explain this observational feature, such as tidal heating (e.g. \citealp{bodenheimer2001}; \citealt{jackson2008}) and ohmic dissipation (e.g. \citealt{Batygin2010}; \citealt{thorngren_fortney2018}), they are not well understood (e.g. \citealt{sarkis2021}). For these reasons, it is very challenging to determine the bulk composition of hot Jupiters based on their current radius and mass. 
Similarly, the strong tidal interactions with the host star, tend to circularize and synchronize their orbits (e.g., \citealt{rasioford1996}; \citealt{guillot1996}), thus hiding important clues about their formation and evolution pathways, such as their primordial orbital eccentricity and stellar obliquity (e.g., \citealt{Albrecht2022}). Thus, detecting WJs and characterizing their orbital properties can help us to distinguish different migration mechanisms, such as angular momentum exchange with the protoplanetary disk (e.g., \citealt{ward1997}) and high-eccentricity migration scenarios (e.g. \citealt{rasioford1996}; \citealt{beauge2012}), both leading to short-period planets ($P \lesssim$ 10 $d$) in nearly circular orbits. 
In this sense, although transiting WJs are rare (mainly because of their low transit probability) and more difficult to characterize than hot Jupiters (due to their longer orbital periods), they provide very valuable information for understanding giant planet formation and their orbital evolution (e.g., \citealt{petrovic2016}). \newline
In this work we present the discovery, confirmation and characterization via ground-based photometric and spectroscopic follow-up of two Jupiter-size transiting substellar objects, detected by the Transiting Exoplanet Survey Satellite (\tess; \citealt{Ricker2015}) mission around TOI-2529 (TIC\,269333648; TYC\,8147-1020-1) and TIC\,4672985 (TYC\,5288-103-1). The paper is structured as follows. In sections \ref{sec:tess_phot} and \ref{sec:ground-based-phot} we present the \tess and ground-based photometric data. In sections \ref{sec:HCI} and \ref{sec:spectroscopy} we describe the ground-based speckle imaging and spectroscopic follow-up, respectively.
The derived stellar parameters are described in section \ref{sec:stellar_par}, the results from the joint fit in section \ref{sec:joint_analysis} and the discussion in section \ref{sec:discussion}. Finally, the summary and conclusions are presented in section \ref{sec:conclusions}.

\section{TESS photometry \label{sec:tess_phot}}

\begin{figure}[!h]
       \includegraphics[angle=0,width = 0.95\columnwidth]{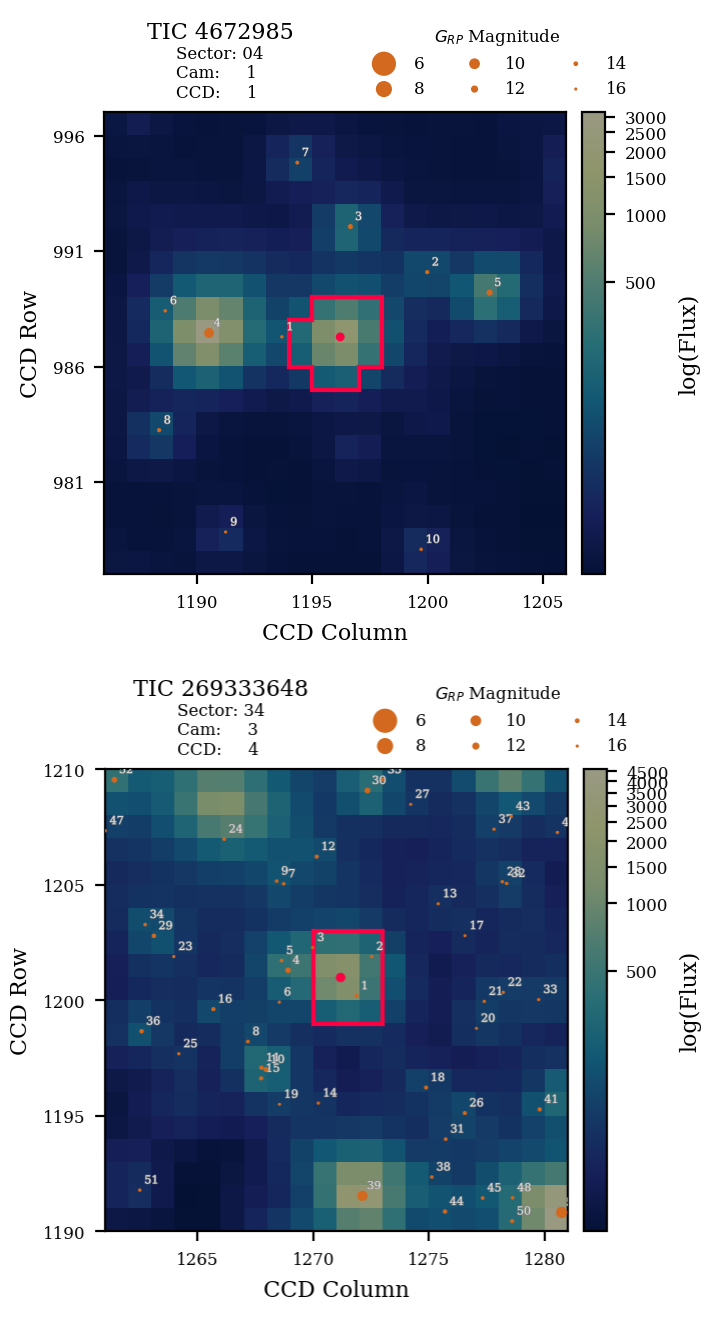}
    \caption{\tess full-frame image of TIC\,4672985, from sector 4 (upper panel) and of TOI-2529 in sector 34 (lower panel). Nearby sources detected by Gaia are also shown. The red box  corresponds to the aperture used for the photometry extraction. \label{fig:FFI_TIC4672985}}
\end{figure}

We identified the transit signal of TIC\,4672985 in the 30-minutes-cadence full-frame image (FFI) data generated by the Science Processing Operations Center at NASA Ames Research Center \citep{jenkinsSPOC2016}, obtained in sector 4 of the \tess primary mission. A query to the Gaia DR3 \citep{GaiaDR3} archive revealed three stars within 42 arcsec (corresponding to two pixels), but all of them are fainter by $\sim$ 7 magnitudes than TIC\,4672985 ($\gtrsim$ 600 times fainter in the G band). No significant contamination from nearby stars is therefore expected. 
For this object, we adopted a simple aperture photometry of two pixels, meaning that only the pixels whose center fall within this radius are included (see Figure \ref{fig:FFI_TIC4672985}). \newline
Similarly, we identified a total of three transit events in 
the TOI-2529 FFI \tess data obtained in sector 8 (30-minute cadence) of the primary mission, and in sectors 34 and 63 of the extended mission, with a cadence of 10 and 2 minutes, respectively. For this star, we used a rectangular aperture to prevent different stars from contaminating the aperture in different sectors (see Figure \ref{fig:FFI_TIC4672985}). However, we found two contaminating stars within the photometry aperture in all three sectors, with G = 15.6 ($\Delta$mag = 4.4) and G = 16.2 mag ($\Delta$mag = 5.0), leading to a combined fractional flux of $\sim$ 3\%. 
These two signals were identified in an effort to characterize warm transiting planets as part of the Warm gIaNts with tEss (WINE) collaboration (\citealt{brahm2019}; \citealt{Brahm2023}). 
The light curves for these two stars in all four TESS sectors were generated with the  \texttt{tesseract\footnote{https://github.com/astrofelipe/tesseract}} pipeline.

\begin{figure}[!h]
       \includegraphics[angle=0,width = \columnwidth]{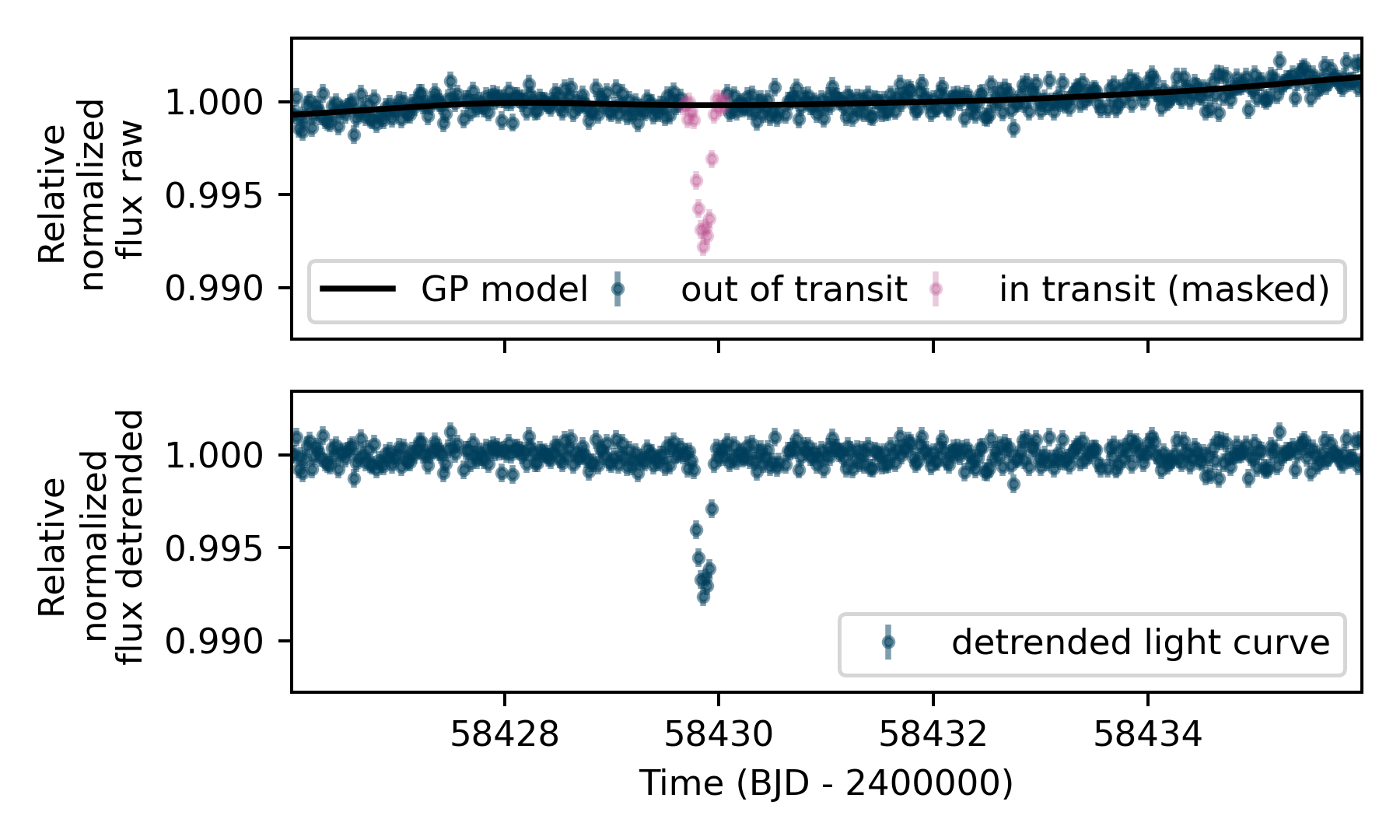}
    \caption{Upper panel: \tess extracted light curve of TIC\,4672985 observed in sector 4. The GP model used to correct for the long-term variability is overplotted (solid black line). Lower panel: Detrended \tess light curve. \label{tic4672_lc_detrend}}
\end{figure}

\begin{figure}[!h]
       \includegraphics[angle=0,width = \columnwidth]{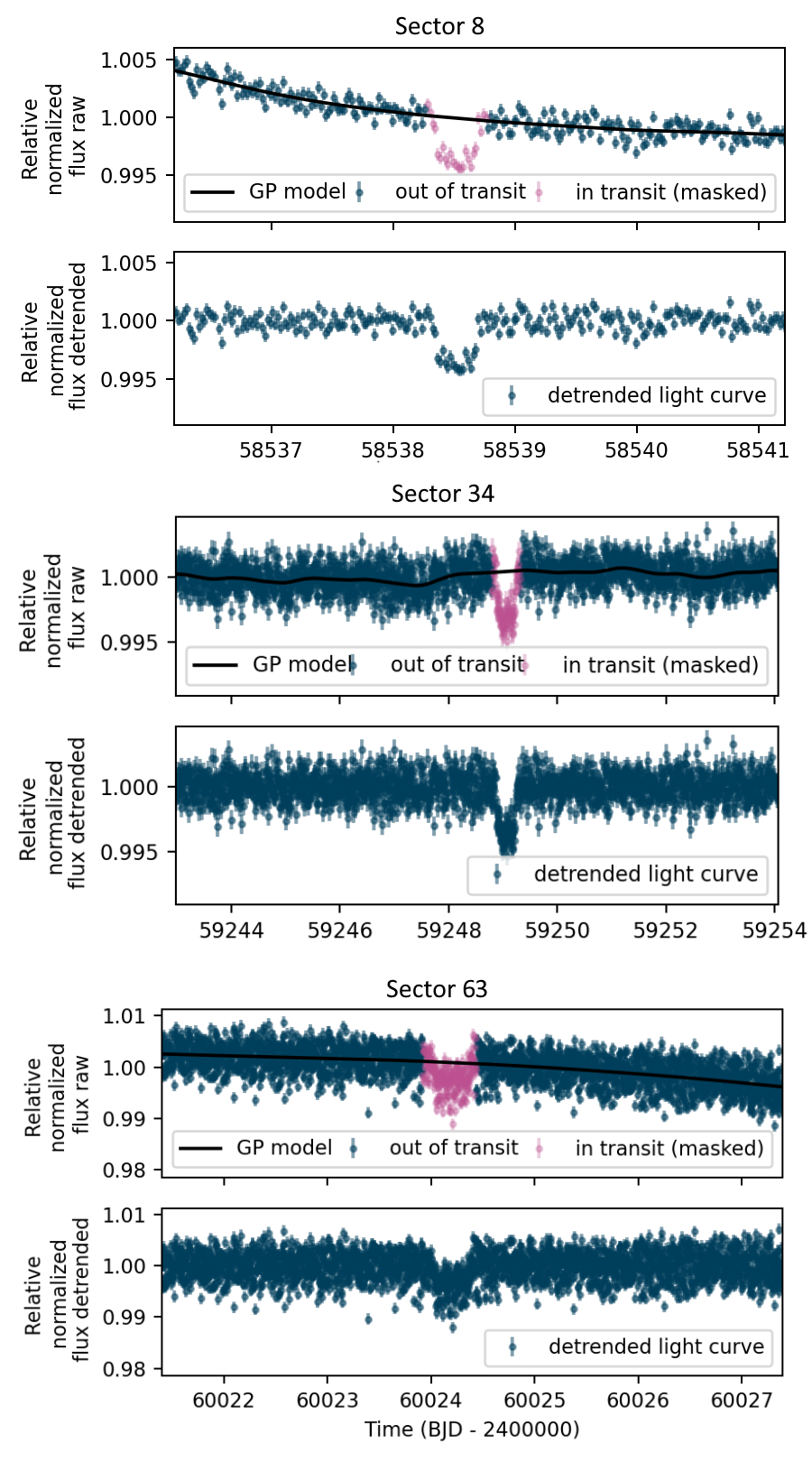}
    \caption{Same as Fig. \ref{tic4672_lc_detrend} but for TOI-2529 in sectors 8, 34 and 63 (upper, middle, and lower panel, respectively).  \label{toi2529_s08}}
\end{figure}

After extracting the light curve, we modeled the long-term continuum variability using a stochastic Gaussian process (GP). To do this, we first masked the transits and then fit the GP using a \celerite\, Matern 3/2 kernel \citep{celerite}. We finally divided the extracted light curve by the resulting GP model to remove this long-term trend. The extracted photometry, GP model, and final detrended light curve for TIC\,4672985 and TOI-2529 are presented in Figures \ref{tic4672_lc_detrend} and \ref{toi2529_s08}. 

\section{Ground-based photometric data \label{sec:ground-based-phot}}

We obtained ground-based photometric follow-up of TIC\,4672985 to confirm the source of the transit detected in the \tess\, light curves, and also to refine the photometric ephemeris, which is particularly important for single-transit events. To do this, we used the available radial velocity data obtained in 2021 (see Sec. \ref{sec:spectroscopy}) in combination with the single \tess transit event in sector 4, to predict future transits that would be observable from the ground. 

\subsection{Callisto}
Callisto is one of the six identical robotic one-meter telescopes of the Search for habitable Planets EClipsing ULtra-cOOl Stars (SPECULOOS; \citealt{Delrez2018}) network. Four of them, including Callisto, are installed at the Paranal observatory in Chile. 
The main scientific goal of SPECULOOS is the detection of
transiting terrestrial planets in orbit around nearby ($<$\,40\,pc) late-type M and L-dwarfs.
Each SPECULOOS telescope is equipped with a 2k $\times$ 2k CCD, 
with a field of view of 12' × 12' and a pixel scale of 0.35 arcsec/pix.
We observed a transit egress of TIC\,4672985\,$b$ with Callisto on 2021 September 8, using the z filter and with a cadence of 40\,s. The extracted normalized photometry is presented in Figure \ref{fig:LC_model_TIC4672985}. 

\subsection{OM-ES}
\begin{figure}[ht]
\includegraphics[width=9cm]{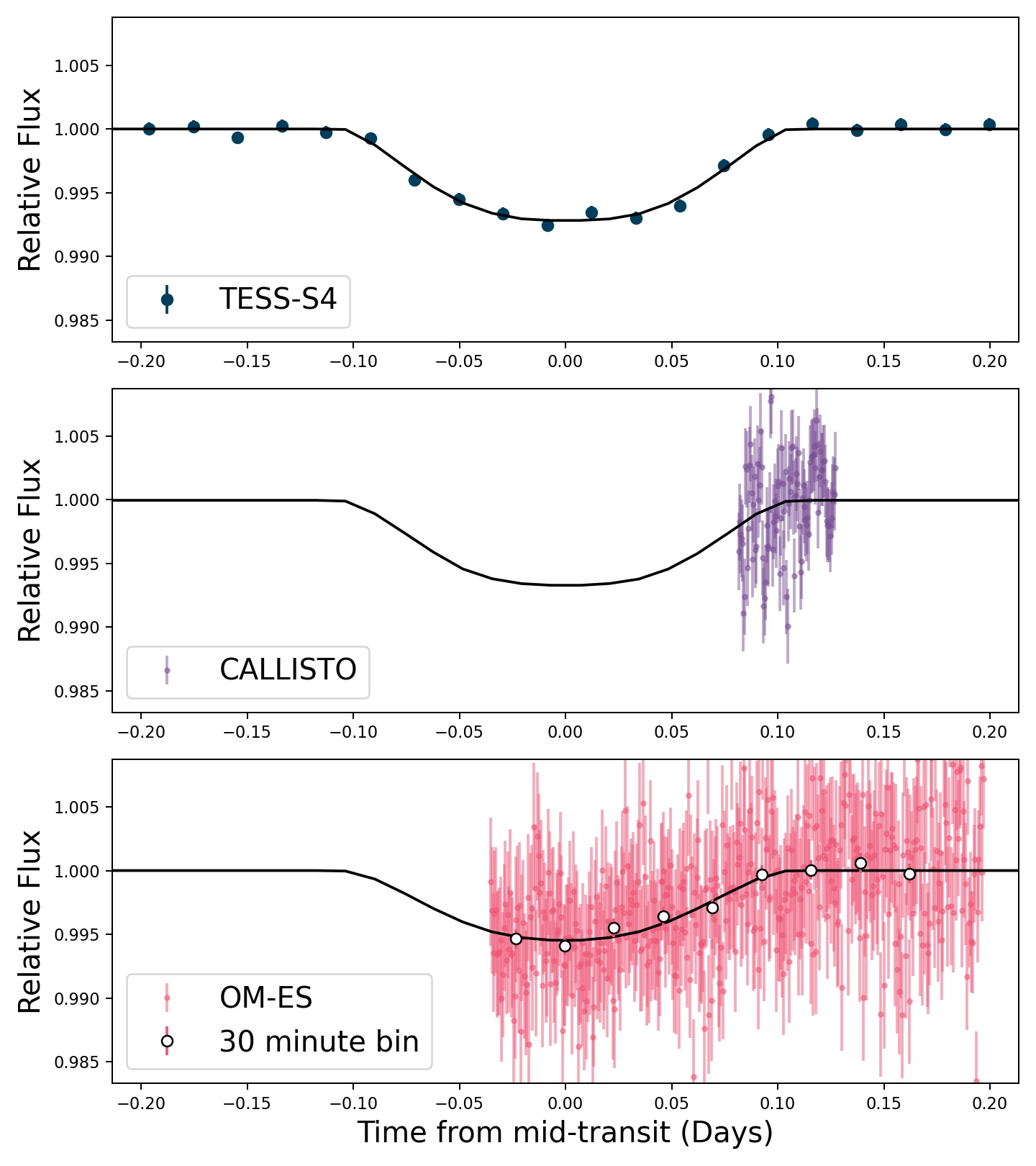}
\caption{Normalized photometric observations of TIC\,4672985 during the transit. The data were obtained in \tess sector 4 (upper panel) and from ground-based telescopes (CALLISTO and OM-ES; middle and lower panels, respectively). The solid lines correspond to the transit models. }
\label{fig:LC_model_TIC4672985}
\end{figure}

The Observatoire Moana -- El Sauce (OM-ES) facility is an equatorial robotic telescope with a 60\,cm aperture, located at El Sauce Observatory, in Chile. OM-ES is equipped with a 2k $\times$ 2k detector, with a pixel scale of of 0.7 arcsec/pix, and has a standard set of Sloan filters u',g',r',i'. Using this telescope, we observed a partial transit of TIC\,4672985 in the night of 2021 November 16. For these observations, we used the i' filter, and we obtained an image every 49\,$s$. The data reduction and photometry extraction was performed with a set of automated routines that were used previously in other facilities (e.g., \citealt{brahm2020}). The extracted normalized photometric data are presented in Figure \ref{fig:LC_model_TIC4672985}.

\section{Speckle imaging \label{sec:HCI}}

\begin{figure}[ht]
\includegraphics[scale=0.33]{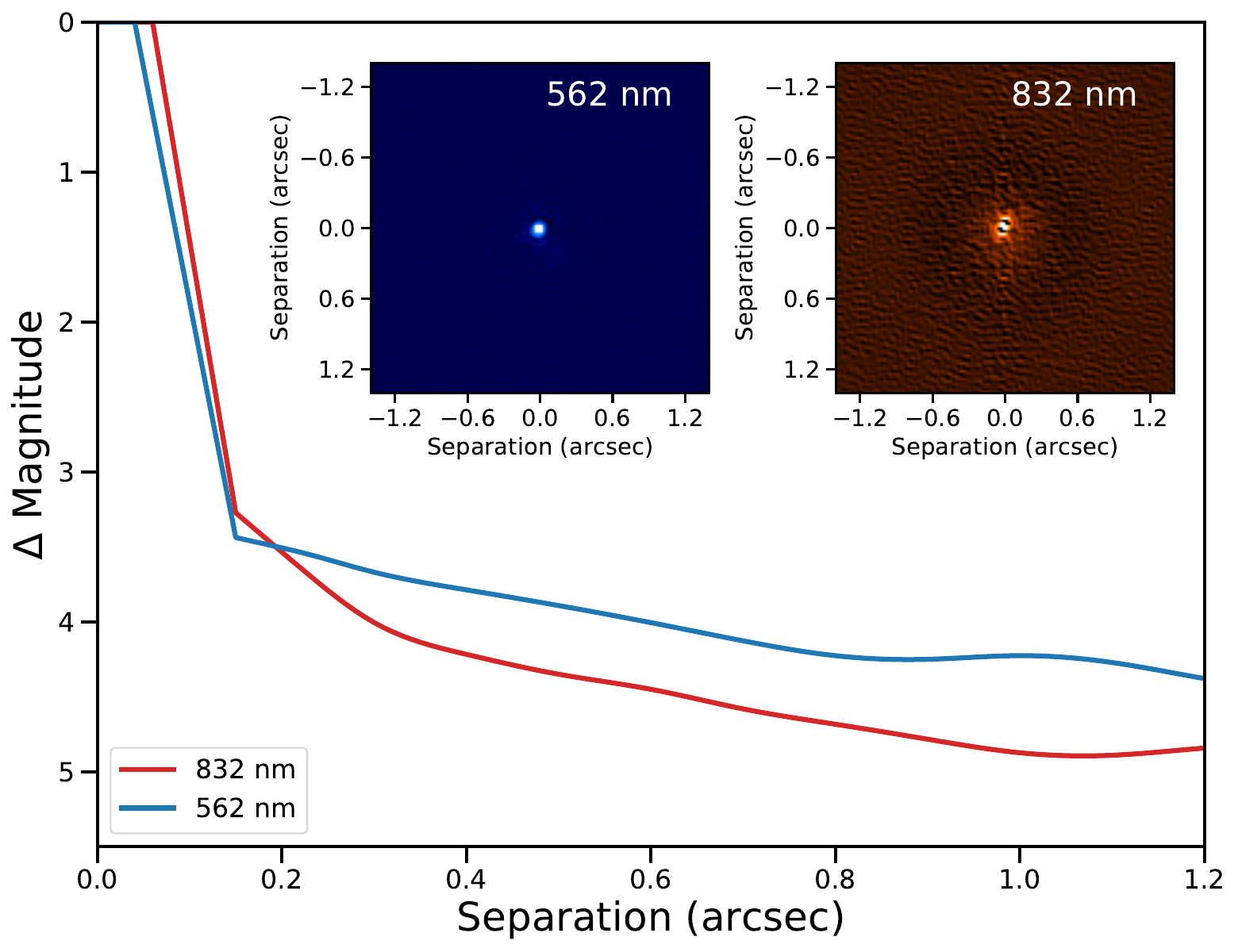}
\caption{Reconstructed NESSI speckle images for
TIC\,4672985. The observations were obtained simultaneously at 532 nm and 832 nm with the blue and red camera (upper left and right images). 
The corresponding 5\,$\sigma$ contrast curves are overplotted (solid red and blue lines, respectively). \label{fig:nessi_HCI}}
\end{figure}

To search for potential sources of blending or flux contamination, we observed TIC\,4672985 with the NN-Explore Exoplanet Stellar Speckle Imager \citep[NESSI;][]{Scott2018} on the WIYN 3.5 m telescope at Kitt Peak National Observatory, on 2019 October 12. We obtained one-minute sequences of short diffraction-limited exposures in each of the 832 nm and 562 nm filters with the red and blue NESSI cameras, respectively. The reconstructed speckle images were created following the procedures described in \citet{Howell2011}. As we show in Figure \ref{fig:nessi_HCI}, the computed $5\sigma$ contrast limits rule out companions brighter than $\Delta$m$_{562}=3.9$ and $\Delta$m$_{832}=4.4$ at separations greater than 0.5 arcsec and $\Delta$m$_{562}=4.2$ and $\Delta$m$_{832}=4.9$ outside of 1 arcsec.

\begin{figure}[ht]
\includegraphics[scale=0.50]{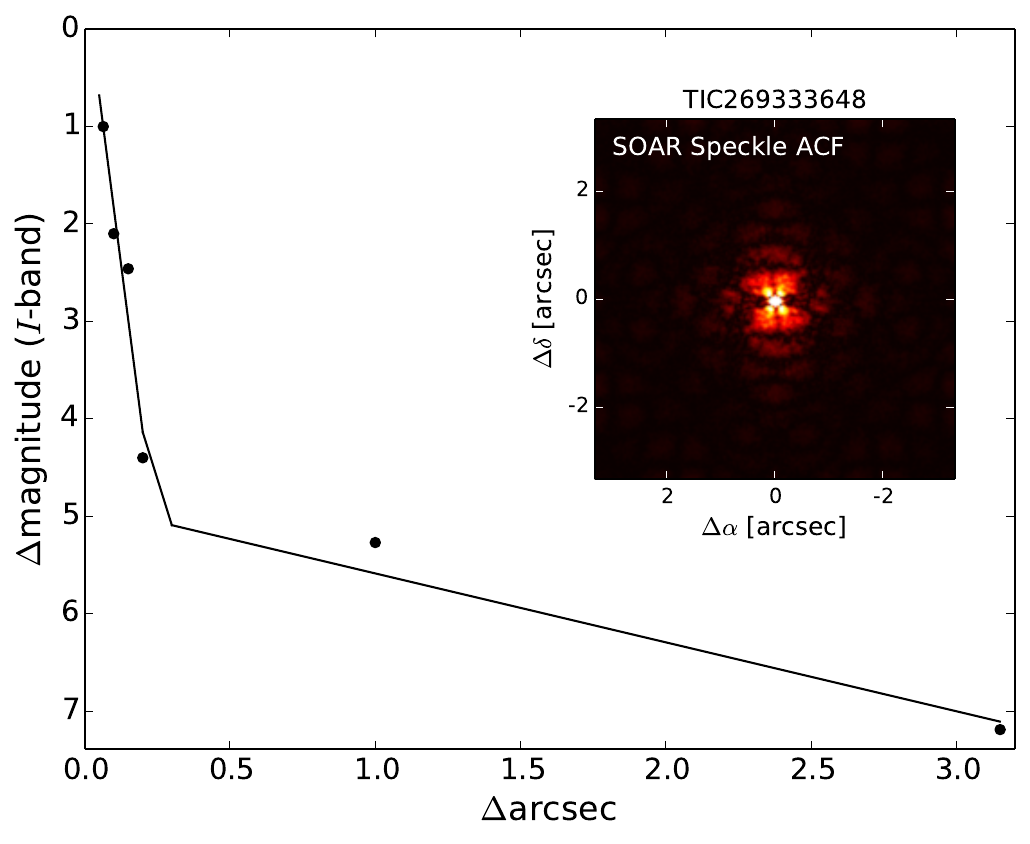}
\caption{Speckle autocorrelation function obtained in the
$I$ band at the SOAR telescope, for TOI-2529. The black dots correspond to the 
5$\sigma$ contrast curve. The solid line corresponds to the linear fit of the data at separations smaller and larger than $\sim$ 0.2 arcsec. \label{fig:SOAR_HCI} }
\end{figure}

Similarly, we searched for stellar companions to TOI-2529 with speckle imaging on the 4.1 m Southern Astrophysical Research (SOAR) telescope \citep{Tokovinin2018}. To do this, we observed this star with the HRCam on 2021 October 1, in the Cousins $I$ band. This observation was sensitive to a star that is fainter by 5.2 magnitudes at an angular distance of 1 arcsec from the target. More details of the observations within the SOAR \tess survey are available in \citet{Ziegler2020}. The 5$\sigma$ detection sensitivity and speckle autocorrelation functions from the observations are shown in Figure \ref{fig:SOAR_HCI}. No nearby stars were detected within 3 arcsec of TOI-2529 in the SOAR observations.

\section{Ground-based spectroscopic follow-up \label{sec:spectroscopy}}

\begin{figure}[ht]
\includegraphics[width=9.5cm]{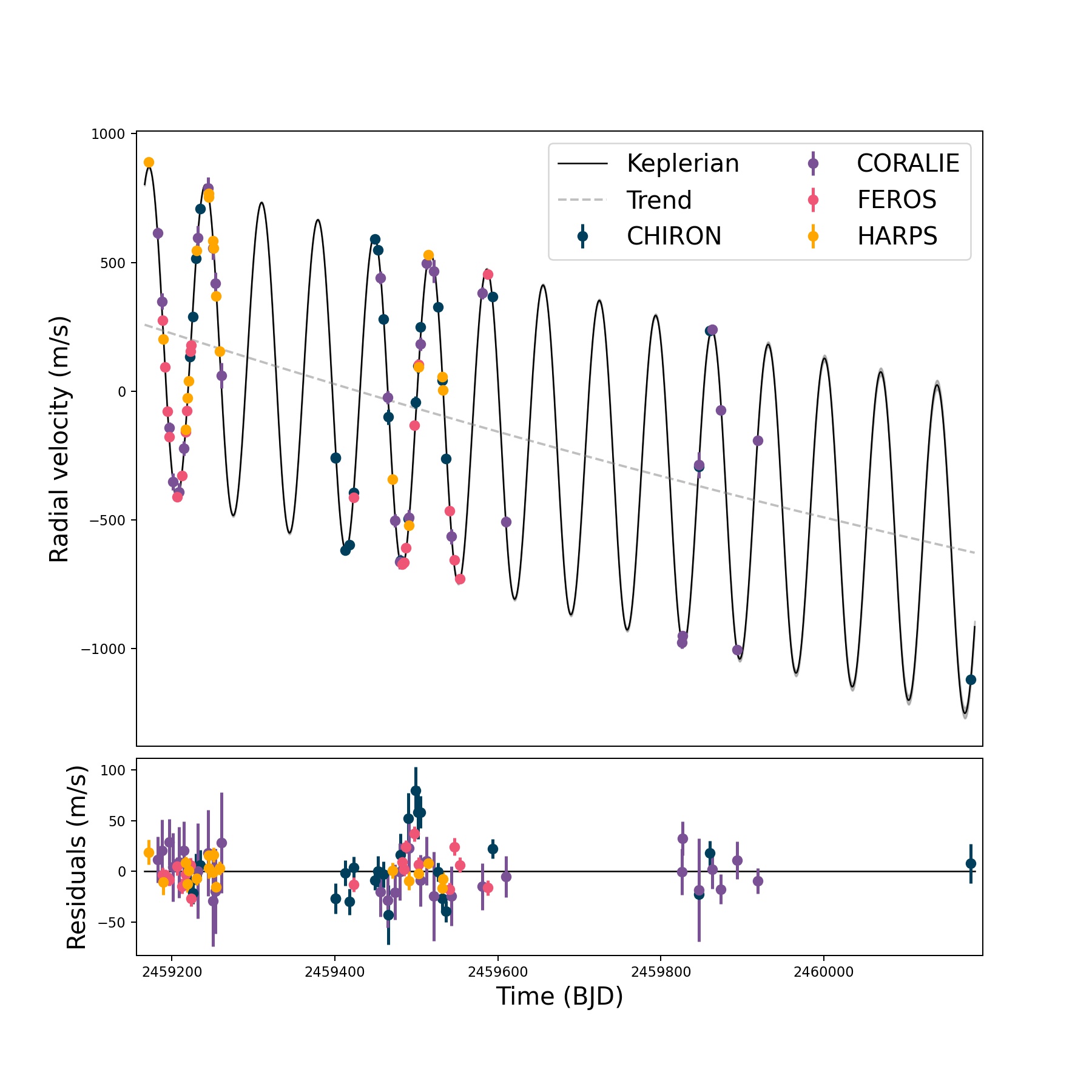}
\caption{RV curve of TIC4672985. The best Keplerian fit and the observed RV trend are overplotted.}
\label{fig:RV_TIC4672985}
\end{figure}

\begin{figure}[ht]
\includegraphics[width=9.5cm]{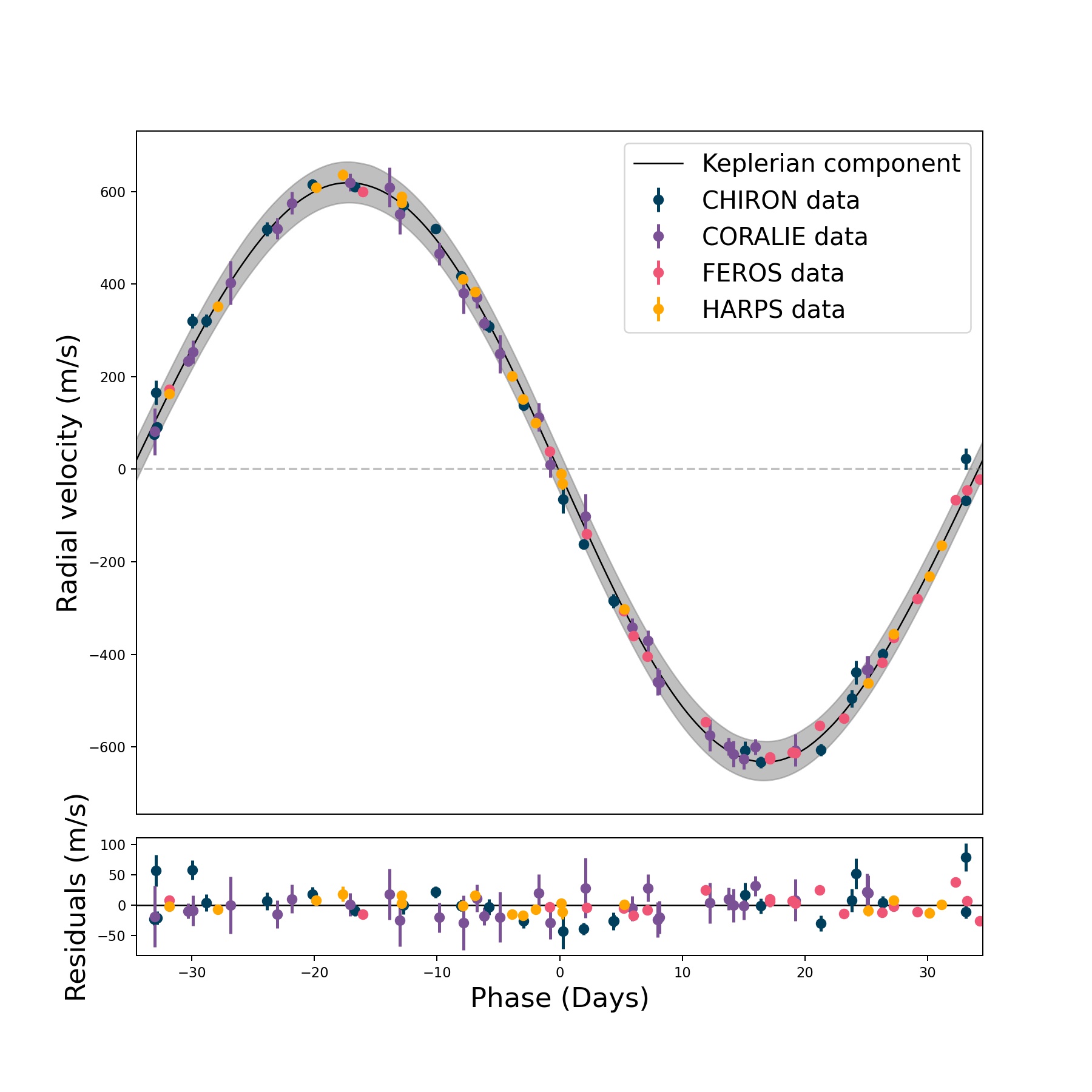}
\caption{Phase-folded RV curve for TIC\,4672985, after subtracting the quadratic trend. The gray region corresponds to 1$\sigma$ uncertainty in the best-fit model.}
\label{fig:RV_TIC4672985_phase}
\end{figure}

\begin{figure}[ht]
\includegraphics[width=9.5cm]{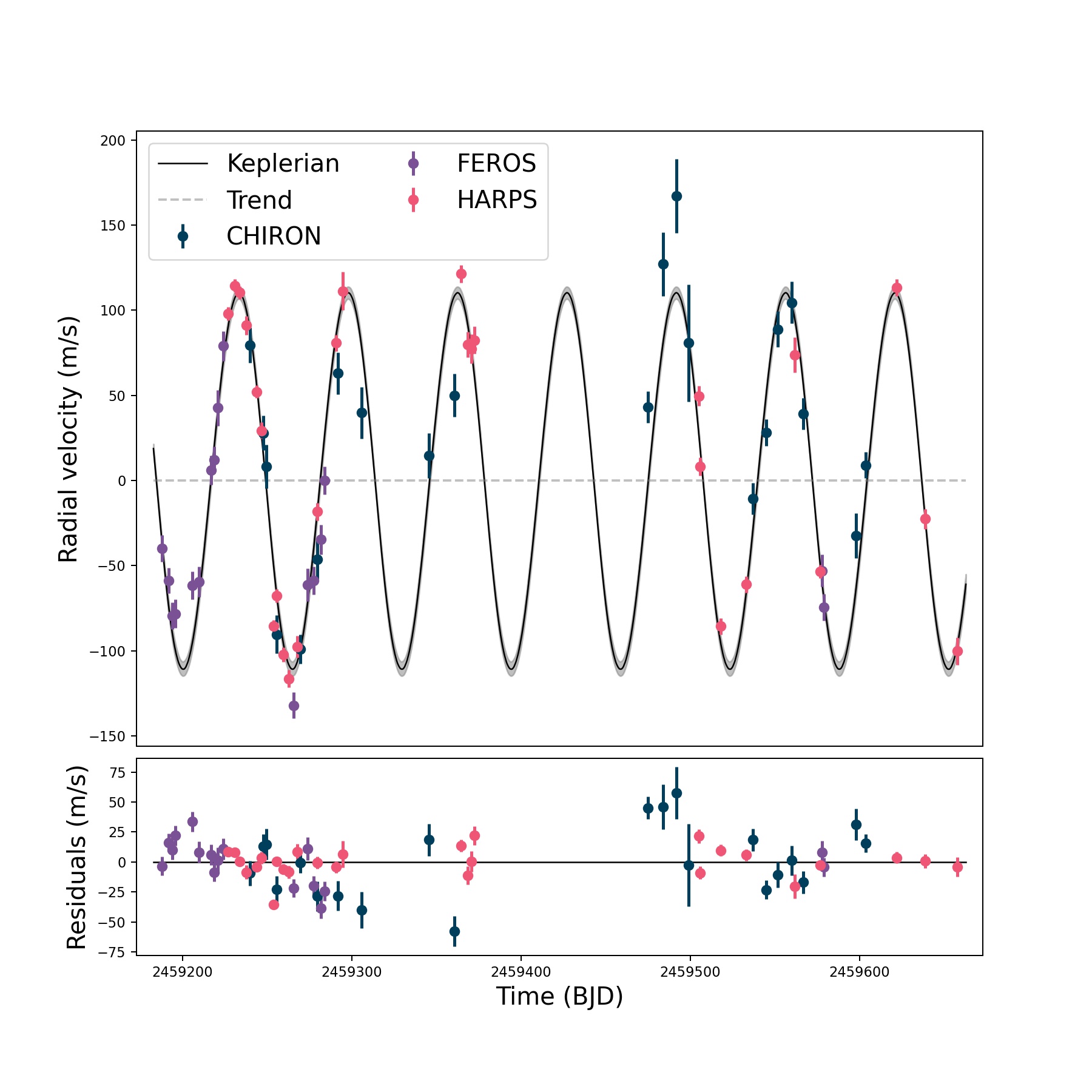}
\caption{Same as Fig. \ref{fig:RV_TIC4672985}, but for TOI-2529. }
\label{fig:RV_TOI2529}
\end{figure}

\begin{figure}[ht]
\includegraphics[width=9.5cm]{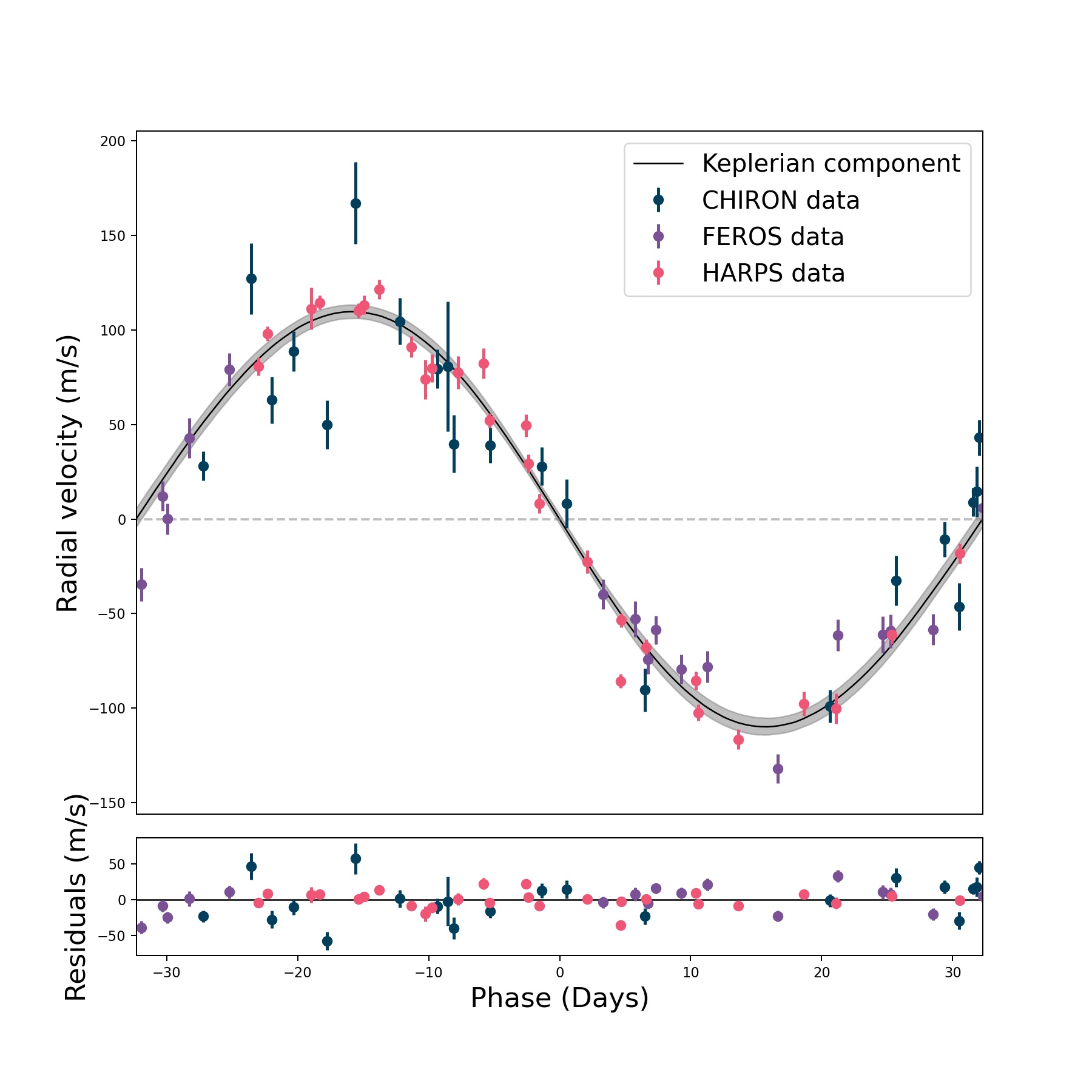}
\caption{Same as Fig. \ref{fig:RV_TIC4672985_phase}, but for TOI-2529.}
\label{fig:RV_TOI2529_phase}
\end{figure}

We performed a spectroscopic follow-up of these two stars, using four different high-resolution spectrographs. From the observed spectra, we computed precision radial velocities (RVs), which were used to confirm the planetary nature of the observed photometric transits and to measure their the masses. Similarly, from the cross-correlation function (CCF), we also measured the line bisectors (BVS; \citealt{Toner1988}). 
The resulting RVs and BVS values, with their corresponding 1$\sigma$ uncertainties are listed in Tables \ref{tic4672_rv} and \ref{tic2693_rv}, and are also presented in Figures \ref{fig:RV_TIC4672985}-\ref{fig:RV_TOI2529_phase}. 

\subsection{CHIRON}

We observed TIC\,4672985 and TOI-2529 with CHIRON \citep{chiron}, as part of a large program in the context of the WINE collaboration. CHIRON is 
a high-resolution fiber-fed spectrograph mounted on the 1.5\,m telescope at the Cerro Tololo Inter-american Observatory (CTIO) in Chile. We obtained a total of 24 spectroscopic epochs for TIC\,4672985 and 21 epochs for TOI-2529. We adopted exposures times of 
1800 s for the two stars, and we used the image-slicer mode, which delivers a spectral resolution of R $\sim$ 80,000. The typical signal-to-noise-ratio (S/N) per extracted pixel was $\sim$ 10 - 20 in the two cases, with corresponding mean RV uncertainties at the $\sim$ 13-15 \ms level.
The data were reduced using the CHIRON pipeline \citep{Paredes2021} and the RVs were computed with the pipeline described in \citet{Jones2019}.

\subsection{FEROS}

We collected a total of 20 individual spectra of TIC\,4672985, between 2020 December 5 and 2022 January 8, with the Fiber-fed Extended Range Optical Spectrograph (FEROS; \citealt{feros}) high-resolution spectrograph ($R\sim48000$), installed at the 2.2\,m telescope, at La Silla. The exposure times were between 900 s and 1800 s. FEROS is equipped with two fibers, allowing for either sky-background subtraction or simultaneous wavelength calibration. The latter is the observing mode we used in our observations. 
Similarly, we observed TOI-2529 a total of 17 times with FEROS between 2020 December 4 and 2021 December 30. For this star the exposure times were between 600 s and 900 s. The data were reduced with the \ceres\, pipeline \citep{ceres}, which also automatically computes the stellar RV and BVS values. The typical uncertainty in the derived FEROS RVs is $\sim$ 7-8 \ms for the two stars.
\subsection{HARPS}

The High Accuracy Radial velocity Planet Searcher (HARPS; \citealt{harps}) is a fibre-fed by the Cassegrain focus of the 3.6\,m telescope at La Silla Observatory. Using this instrument, we obtained 27 spectra for  TOI-2529 and 18 for TIC\,4672985. We selected the high-resolution mode (1 arcsec fibre, leads to R $\sim$ 115,000) and the simultaneous wavelength calibration for our observations. 
The exposure time was 900\,s for TIC\,4672985 and between 600\,s and 900\,s for TOI-2529. As for FEROS, the data reduction and RV computation was made with \ceres. The mean RV uncertainty is $\sim$ 6 \ms and 8 \ms, for TIC\,4672985 and TOI-2529, respectively. 

\subsection{CORALIE}
We collected a total of 29 spectra of TIC\,4672985 using CORALIE \citep{coralie}, which is mounted on the 1.2\,m Leonhard Euler Telescope at La Silla observatory.
CORALIE is an echelle high-resolution spectrograph that delivers a resolving power of R $\sim$ 60,000. This instrument is also equipped with two optical fibres, the second of which is fed by either a sky or a simultaneous calibration source, in this case, a Fabry-Perot interferometer. As for FEROS and HARPS, we used the latter configuration to correct for the night drift. The data reduction and RV computation was performed with the CORALIE data reduction software.

\section{Stellar parameters  \label{sec:stellar_par}}

We derived the stellar properties of the two host stars, using the zonal atmospheric stellar parameters estimator (\zaspe; \citealt{zaspe}) code, following an iterative procedure described in detail elsewhere (e.g., \citealt{brahm2019}; \citealt{Hobson2021}). 
Briefly, we first coadded the FEROS spectra to create a high S/N observed template of each star. We then used \zaspe to compare the resulting template with a grid of synthetic spectra from the ATLAS9 models \citep{castelli_kurucz2003}. After exploring a wide parameter range, we derived a first set of stellar atmospheric parameters (\teff, log\,$g$, [Fe/H] and \vsini). 
Then, the physical stellar parameters (\mstar, \rstar, \lstar, $\rho_\star$) and the stellar age and visual extinction in the line of sight (A$_v$)  were obtained by comparing the stellar broadband photometry, which was converted into absolute magnitude using the GAIA DR2 \citep{gaia:dr2} parallaxes, with the synthetic magnitudes from the \parsec stellar evolutionary models \citep{bressan:2012}. 
In this process we fixed the metallicity found in the first step and used the initial \teff\,as a prior for this second step. We repeated this procedure in an iterative way until convergence was reached. The resulting atmospheric and physical parameters for the two stars are listed in Table \ref{Tab:atm_par}. We note that for the rest of the paper, we adopt the \zaspe stellar parameters. 
For comparison, we also computed atmospheric and physical parameters for the two stars using the \SPECIES code \citep{Soto2021}. They are also listed in Table \ref{Tab:atm_par}. In general, the two methods 
agree very well. We note that the uncertainties listed in Table \ref{Tab:atm_par} correspond to the internal uncertainties obtained by the different methods we used, and they do not consider systematic errors. 
However, following \citet{Tayar2022}, we add systematic uncertainties of 2\% in quadrature to the stellar luminosity, an uncertainty of 4\% to the stellar radius, an uncertainty of 5\% to the stellar mass, and an uncertainty of 20\% to the stellar age in the following sections. This significantly affects the final planet parameters and their corresponding uncertainties. \newline
Finally, we performed a periodogram analysis of the V-band all sky automated survey (ASAS; \citealt{ASAS1997}) and the all-sky automated survey for supernovae
(ASAS-SN; \citealt{Kochanek2017}) photometric data of these two stars to estimate the stellar age via gyrochronology. No significant periodicity was found in the periodogram of either star (see Fig. \ref{toi2529_photometry_periodogram} and \ref{tic4672_photometry_periodogram}), and their rotational period could therefore not be derived.

\begin{table}
\centering
\caption{Stellar parameters. \label{Tab:atm_par}}
\begin{tabular}{lrrr}
\hline\hline
\vspace{-0.3cm} \\
Parameter & TOI-2529 & TIC\,4672985 & Source \\
\hline \vspace{-0.3cm}                                                        \\
\vspace{0.1cm}
 \teff   [K]     & 5802$_{-52}^{+60}$         & 5757$_{-65}^{+72}$       & \zaspe     \\  \vspace{0.1cm}
                 & 5793$_{-50}^{+50}$         & 5754$_{-66}^{+66}$       & \SPECIES    \\ \vspace{0.1cm}
 log\,$g$ [dex]  & 4.03$_{-0.01}^{+0.01}$     & 4.32$_{-0.02}^{+0.02}$   & \zaspe  \\ \vspace{0.1cm}
                 & 4.05$_{-0.08}^{+0.08}$     & 4.32$_{-0.08}^{+0.08}$   & \SPECIES  \\ \vspace{0.1cm}
 \feh [dex]      & 0.10$_{-0.03}^{+0.03}$     & 0.14$_{-0.05}^{+0.05}$   & \zaspe   \\ \vspace{0.1cm}
                 & 0.13$_{-0.05}^{+0.05}$     & 0.20$_{-0.04}^{+0.04}$   & \SPECIES   \\ \vspace{0.1cm}
 \mstar [\msun]  & 1.11$^{+0.01}_{-0.02}$     & 1.01$_{-0.03}^{+0.03}$   & \zaspe   \\ \vspace{0.1cm}
                 & 1.13$^{+0.02}_{-0.02}$     & 1.04$^{+0.02}_{-0.02}$   & \SPECIES   \\ \vspace{0.1cm}
 \rstar [\rsun]  & 1.70$_{-0.03}^{+0.02}$     & 1.15$_{-0.01}^{+0.01}$   & \zaspe   \\ \vspace{0.1cm}
                 & 1.69$_{-0.03}^{+0.03}$     & 1.15$_{-0.02}^{+0.02}$   & \SPECIES   \\ \vspace{0.1cm}
 \lstar  [\lsun] & 2.91$_{-0.12}^{+0.15}$     & 1.30$_{-0.05}^{+0.06}$   & \zaspe \\ \vspace{0.1cm}
                 & 2.88$^{+0.12}_{-0.11}$     & 1.26$^{+0.04}_{-0.04}$   & \SPECIES   \\ \vspace{0.1cm}
  Age    [Gyr]   & 7.23$_{-0.55}^{+0.41}$     & 7.71 $_{-1.46}^{+1.45}$  & \zaspe \\ \vspace{0.1cm}
                 & 7.05$^{+0.40}_{-0.34}$     & 6.82$^{+1.02}_{-0.98}$   & \SPECIES   \\ \vspace{0.1cm}
  A$_V$  [mag]   & 0.45$_{-0.06}^{+0.08}$     & 0.11 $_{-0.06}^{+0.07}$  & \zaspe \\ \vspace{0.1cm}
                 & 0.42$_{-0.10}^{+0.10}$     & 0.07$_{-0.10}^{+0.10}$   & \SPECIES   \\ \vspace{0.1cm}
$\varv$\,sin\,$i$ [k\ms]  & 3.7$_{-0.3}^{+0.3}$      &  3.5$_{-0.4}^{+0.4}$     & \zaspe     \\ \vspace{0.1cm}
T [mag]          & 10.669 $\pm$ 0.006         &   10.992 $\pm$ 0.007     &  TICv8$^a$ \\ \vspace{0.1cm}
B [mag]          & 12.12 $\pm$ 0.16           &   \dots                  &  Tycho-2$^b$ \\ \vspace{0.1cm}
                 & \dots                      &   12.275 $\pm$ 0.023     &  APASS$^c$ \\ \vspace{0.1cm}
V [mag]          & 11.53 $\pm$ 0.13           &   \dots                  &  Tycho-2 \\ \vspace{0.1cm}
                 & \dots                      &   11.584 $\pm$ 0.014     &  APASS \\ \vspace{0.1cm}
G [mag]          & 11.190 $\pm$ 0.003         &   11.4391 $\pm$ 0.003    &  \gaia$^d$   \\ \vspace{0.1cm}
J [mag]          & 9.935 $\pm$ 0.024          &   10.354 $\pm$ 0.023     &  2MASS$^e$  \\ \vspace{0.1cm}
H [mag]          & 9.620 $\pm$ 0.026          &   10.059 $\pm$ 0.022     &  2MASS  \\ \vspace{0.1cm}
K [mag]          & 9.489 $\pm$ 0.024          &   9.97 $\pm$ 0.023       &  2MASS  \\    \vspace{0.1cm}
\vspace{-0.2cm} \\
\hline\hline
\end{tabular}
\tablebib{
$(a):$\,\citealt{stassun2019}; $(b):$\,\citealt{Hog2000}; $(c): $\,\citealt{Munari2014}; $(d):$\,\citealt{GaiaEDR32020}; $(e):$\,\citealt{Cutri2003}}
\end{table}

\section{Joint photometric and radial velocity analysis \label{sec:joint_analysis}}

\begin{figure}[ht]
\includegraphics[width=9.0cm]{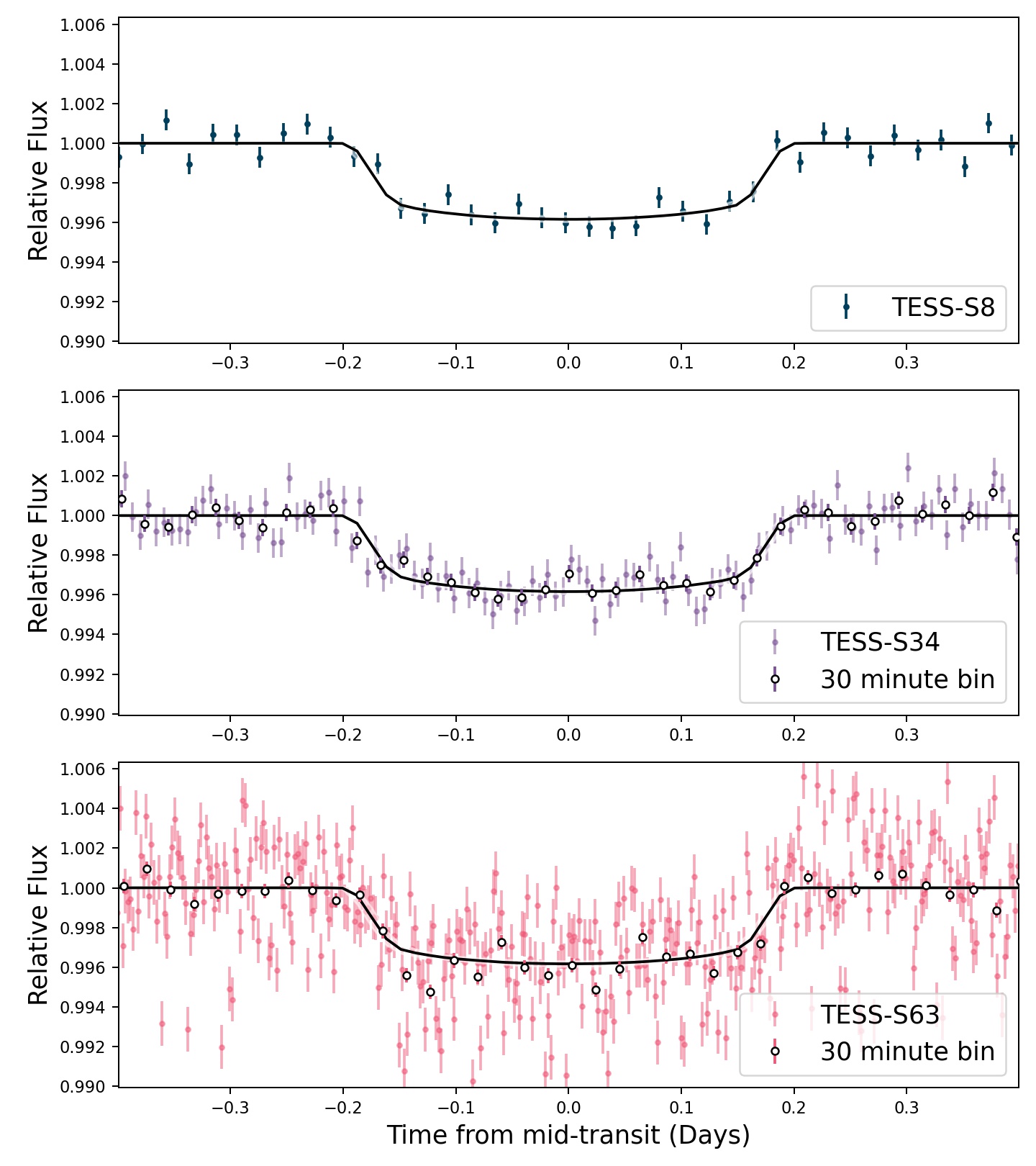}
\caption{Normalized \tess photometry of TOI-2529 around the transit in sectors 8, 34, and 63 (upper, middle, and lower panels, respectively). The 30-minute binned data and the best-fit transit model are overplotted.}
\label{fig:LC_model_TOI2529}
\end{figure}

We jointly modeled the ground- and space-based photometry with the derived RVs for the two systems presented here. To do this we used 
\juliet \citep{juliet}, a python-based package that employs Bayesian inference and nested sampling to achieve model fitting and comparison. 
\juliet uses the \radvel package \citep{Fulton2018} for the RV data modeling, and \batman \citep{Kreidberg2015} for the photometric transit model. In addition, we used the \dynesty package \citep{Speagle2020}, which is incorporated in \juliet, to sample the posterior distributions. 
We used uniform priors for the orbital and transit parameters.
Based on the effective temperature of both host stars and following \citep{Espinoza2016}, we adopted a logarithmic limb-darkening law for both ground- and space-based photometry, also setting uniform priors on the $q_1$ and $q_2$ parameters (see \citealt{kipping2013}). In the case of TIC\,4672985, we fixed the dilution factor to 1, while in the case of TOI-2529, we fit this parameter by adopting a uniform prior based on the expected level of contamination by nearby sources (see Sec. \ref{sec:tess_phot}).
Following the approach presented in \citet{juliet}, we used the estimated stellar density to constrain $a/R_*$, this time using a normal distribution prior, with $\mu$ and $\sigma$ derived using \zaspe. 
Finally, we adopted uniform priors for the RV zeropoint and extra jitter, while for the photometric extra jitter, we used log-uniform priors.
We note that for TIC\,4672985, we also modeled the RV trend with a linear and quadratic trend. The resulting Bayesian evidence favored the quadratic model ($\Delta logz$ = 6.1). 
The prior and resulting posterior distributions of the pertinent parameters for TIC\,4672985\,$b$ and TOI-2529\,$b$ are presented in Tables \ref{tab:TIC4672985_mod_par} and \ref{tab:TOI-2529_mod_par}, respectively\footnote{For the equilibrium temperature we adopted a zero Bond albedo and $\beta$ = 1; see \citet{Kaltenegger2011}}. Similarly, the transit models for TIC\,4672985\,$b$ and TOI-2529\,$b$ are presented in Figures \ref{fig:LC_model_TIC4672985} and \ref{fig:LC_model_TOI2529}, respectively.
Finally, for these two systems, no significant periodic signal is detected in the RV residuals {\bf (see Figures \ref{tic4627_periodogram} and \ref{toi2529_periodogram})}.


\begin{table}\caption{TIC\,4672985 Model Parameters}              
\label{tab:TIC4672985_mod_par}      
\centering                                      
\begin{tabular}{l c r}          
\hline\hline                        
Parameter & Prior & Value \\    
\hline                                   
    $P$ (days)  & $\mathcal{U}$(68.9,69.1) & $69.0480\substack{+0.0004 \\ -0.0005}$\\
    $T_0$ (BJD - 2450000) & $\mathcal{U}$(8429.7,8429.8) & $8429.8577\substack{+0.0018 \\ -0.0016}$\\
    $e$ & $\mathcal{U}$(0.0,0.4) & $0.018\substack{+0.004 \\ -0.004}$\\
    $\omega$ & $\mathcal{U}$(0.0,360.0) & $126.3\substack{+11.3 \\ -7.6}$\\
    $R_p/R_*$ & $\mathcal{U}$(0.01,0.15) & $0.089\substack{+0.005 \\ -0.004}$\\
    $b$ & $\mathcal{U}$(0.0,1.0) & $0.90\substack{+0.01\\ -0.01}$\\
    $\rho_*$ (kg\,m$^{-3}$) & $\mathcal{N}$(1023.0,57.999) & $939.0\substack{+99.8 \\ -101.8}$\\
    $K$ (m\,s$^{-1})$ & $\mathcal{U}$(500.0,700.0) & $625.4\substack{+3.3 \\ -3.4}$\\
    $\mu^{coralie}$ (m\,s$^{-1})$& $\mathcal{U}$(-500.0,500.0) & $60.7\substack{+6.6 \\ -6.2}$\\
    $\sigma^{coralie}_w$ (m\,s$^{-1}$) & $\mathcal{U}$(5.0,40.0) & $9.1\substack{+5.4 \\ -2.8}$\\
    $\mu^{chiron}$ (m\,s$^{-1})$& $\mathcal{U}$(-500.0,500.0) & $28.7\substack{+6.1 \\ -6.0}$\\
    $\sigma^{chiron}_w$ (m\,s$^{-1}$) & $\mathcal{U}$(0.0,50.0) & $24.1\substack{+5.9 \\ -4.8}$\\
    $\mu^{harps}$ (m\,s$^{-1})$& $\mathcal{U}$(-500.0,500.0) & $-250.4\substack{+3.5 \\ -3,2}$\\
    $\sigma^{harps}_w$ (m\,s$^{-1}$) & $\mathcal{U}$(0.0,15.0) & $9.5\substack{+2.8 \\ -2.6}$\\
    $\mu^{feros}$ (m\,s$^{-1})$& $\mathcal{U}$(-500.0,500.0) & $215.2\substack{+4.1 \\ -4.3}$\\
    $\sigma^{feros}_w$ (m\,s$^{-1}$) & $\mathcal{U}$(0.0,25.0) & $15.4\substack{+3.4 \\ -3.1}$\\
    $q_1^{tess-S4}$  & $\mathcal{U}$(0.0,1.0) & $0.53\substack{+0.32 \\ -0.33}$\\
    $q_1^{OM-ES}$  & $\mathcal{U}$(0.0,1.0) & $0.61\substack{+0.27 \\ -0.28}$\\
    $q_1^{callisto}$ & $\mathcal{U}$(0.0,1.0) & $0.47\substack{+0.36 \\ -0.31}$\\
    $q_2^{tess-S4}$  & $\mathcal{U}$(0.0,1.0) & $0.74\substack{+0.17 \\ -0.23}$\\
    $q_2^{OM-ES}$ & $\mathcal{U}$(0.0,1.0) & $0.12\substack{+0.14 \\ -0.09}$\\
    $q_2^{callisto}$ &$\mathcal{U}$(0.0,1.0) & $0.51\substack{+0.30 \\ -0.31}$\\
    $\sigma_w^{tess-S4}$ (ppm)& $\mathcal{J}$(0.01,1000.0) & $1.0\substack{+18.8 \\ -0.91}$\\
    $\sigma_w^{OM-ES}$ (ppm) & $\mathcal{J}$(0.01,10000.) & $3.8\substack{+135.2 \\ -3.7}$\\
    $\sigma_w^{callisto}$ (ppm) &$\mathcal{J}$(0.01,10000.) & $1818.2\substack{+562.0 \\ -1385.4}$\\
    $rv_{slope}$ (\ms\,d$^{-1}$) & $\mathcal{U}$(-10.0,10.0) & $-0.94\substack{+0.01 \\ -0.01}$\\
    $rv_{quad}$ (\ms\,d$^{-2}$)& $\mathcal{U}$(-0.1,0.1) & $0.00015\substack{+0.00004 \\ -0.00004}$\\
\hline                                   
    $a$ (AU) & & $0.330\substack{+0.019 \\ -0.019}$\\
    $R_p$ (R$_{\rm J}$) & & $1.026\substack{+0.067 \\ -0.065}$\\
    $M_p$ (M$_{\rm J}$) & & $12.74\substack{+1.01 \\ -1.01}$\\
    $T_{eq}$ (K) & & $517.2\substack{+11.3 \\ -11.2}$\\
\hline                                             
\end{tabular}
\footnotesize{Note: The instrumental zeropoints were computed after subtracting the RV mean value.}\\
\end{table}

\begin{table}\caption{TOI-2529 model parameters}              
\label{tab:TOI-2529_mod_par}      
\centering                                      
\begin{tabular}{l c r}          
\hline\hline                        
Parameter & Prior & Value \\    
\hline                                   
    $P$ (days) & $\mathcal{U}$(64.5,64.6) & $64.5949\substack{+0.0003 \\ -0.0003}$\\
    $T_0$ (BJD - 2450000) & $\mathcal{U}$(8538.5,8538.6) & $8538.5151\substack{+0.0046 \\ -0.0041}$\\
    $e$ & $\mathcal{U}$(0.0,0.5) & $0.021\substack{+0.024 \\ -0.015}$\\
    $\omega$ & $\mathcal{U}$(0.0,360.0) & $103.8\substack{+106.5 \\ -39.3}$\\
    $R_p/R_*$ & $\mathcal{U}$(0.01,0.15) & $0.061\substack{+0.001 \\ -0.001}$\\
    $b$ & $\mathcal{U}$(0.0,1.0) & $0.68\substack{+0.03 \\ -0.04}$\\
    $\rho_*$ (kg\,m$^{-3}$) & N(328.7,41.3) & $326.0\substack{+40.7 \\ -41.9}$\\
    $K$ (m\,s$^{-1})$ & $\mathcal{U}$(50.0,200.0) & $110.5\substack{+3.1 \\ -3.2}$\\
    $\mu^{harps}$ (m\,s$^{-1})$& $\mathcal{U}$(-100.0,100.0) & $-17.8\substack{+2.6 \\ -2.6}$\\
    $\sigma^{harps}_w$ (m\,s$^{-1}$) & $\mathcal{U}$(0.0,15.0) & $11.2\substack{+1.9 \\ -1.6}$\\
    $\mu^{feros}$ (m\,s$^{-1})$& $\mathcal{U}$(-100.0,100.0) & $38.3\substack{+4.7 \\ -5.2}$\\
    $\sigma^{feros}_w$ (m\,s$^{-1}$) & $\mathcal{U}$(0.0,25.0) & $17.3\substack{+2.6 \\ -2.6}$\\
    $\mu^{chiron}$ (m\,s$^{-1})$& $\mathcal{U}$(-100.0,100.0) & $-32.9\substack{+6.3 \\ -6.2}$\\
    $\sigma^{chiron}_w$ (m\,s$^{-1}$) & $\mathcal{U}$(10.0,40.0) & $27.2\substack{+5.5 \\ -5.2}$\\
    $\sigma_w^{tess}-S8$ (ppm)  & $\mathcal{J}$(0.01,10000.) & $444.8\substack{+49.9 \\ -42.5}$\\
    $\sigma_w^{tess}-S34$ (ppm) & $\mathcal{J}$(0.01,10000.) & $673.2\substack{+25.0 \\ -27.0}$\\
    $\sigma_w^{tess}-S63$ (ppm)& $\mathcal{J}$(0.01,10000.) & $2362.2\substack{+42.2 \\ -38.6}$\\
    $q_1^{tess}$          & $\mathcal{U}$(0.0,1.0) & $0.38\substack{+0.38 \\ -0.28}$\\
    $q_2^{tess}$          & $\mathcal{U}$(0.0,1.0) & $0.52\substack{+0.27 \\ -0.29}$\\
    $m_{dilution}^{tess}$ & $\mathcal{U}$(0.96,0.98) & $0.97\substack{+0.01 \\ -0.01}$\\ \vspace{-0.3cm}\\
\hline                                   
    $a$ (AU) & & $0.327\substack{+0.020 \\ -0.020}$\\
    $R_p$ (R$_{\rm J}$) & & $1.030\substack{+0.050 \\ -0.050}$\\
    $M_p$ (M$_{\rm J}$) & & $2.340\substack{+0.197 \\ -0.195}$\\
    $T_{eq}$ (K) & & $636.0\substack{+15.6 \\ -15.7}$\\
\hline            
\hline
\end{tabular}
\footnotesize{Note: The instrumental zeropoints were computed after subtracting the RV mean value.}\\
\end{table}

\section{Discussion \label{sec:discussion}}

\begin{figure}
    \centering
    \includegraphics[angle=90,width=\linewidth]{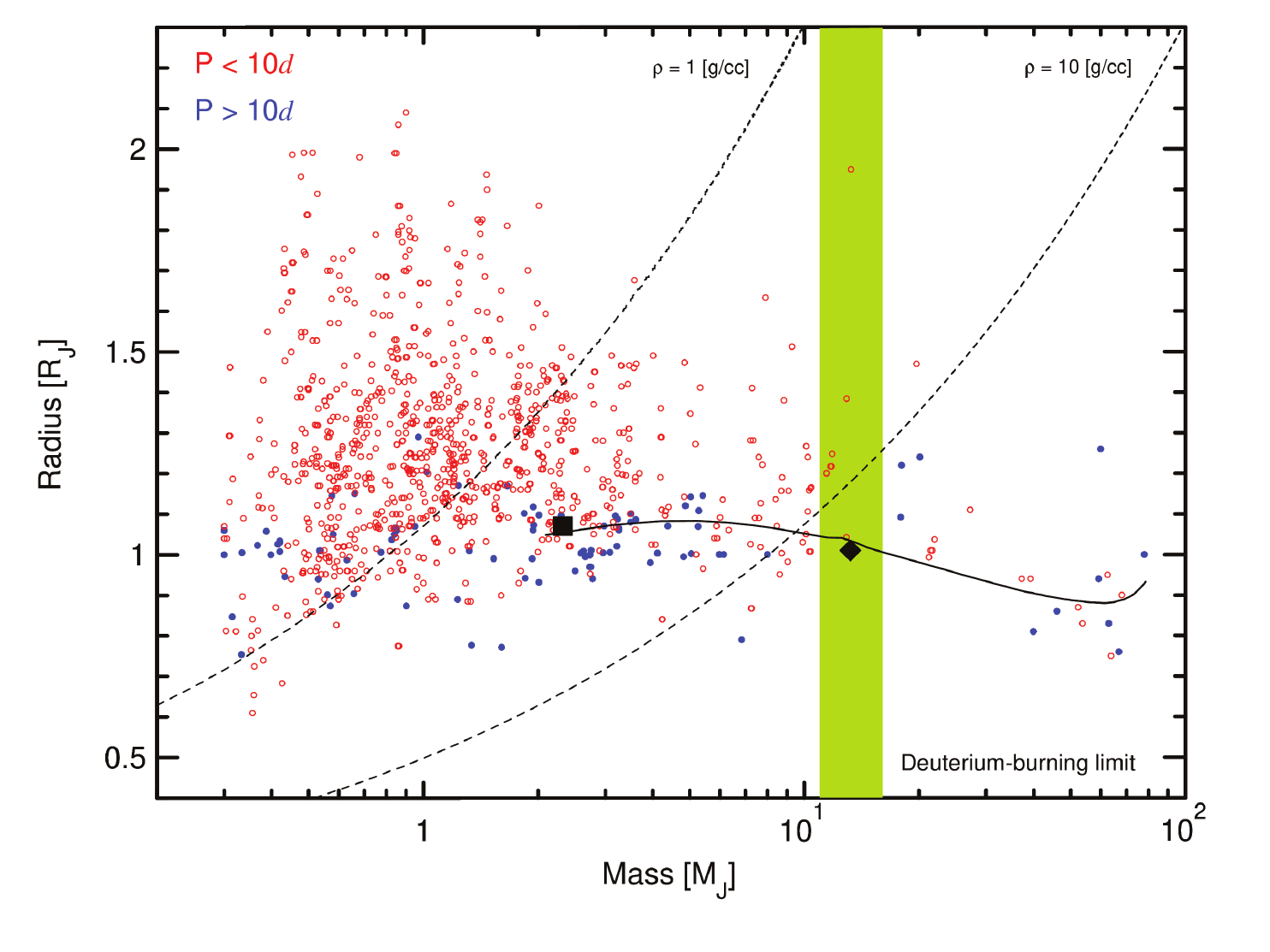}
    \caption{Mass-radius distribution of known transiting giant planets as of June 30, 2023. The open red circles and blue dots correspond to orbital periods shorter and longer than 10 $d$, respectively. The positions of TIC\,4672985\,$b$ (black diamond) and TOI-2529\,$b$ (black square) are also shown. The shaded area represents the theoretical deuterium-burning limit. The solid line corresponds to a 1 Gyr old and solar metallicity ATMO2020 isochrone. Two isodensity curves for 1 and 10 [gr\,cm$^{-3}$] are also plotted (dashed left and right lines, respectively).  }
    \label{fig:mass_rad}
\end{figure}

\subsection{TOI-2529\,$b$ and TIC\,4672985\,$b$ in the context of transiting giant planets}

\begin{figure}
    \centering
    \includegraphics[angle=90,width=\linewidth]{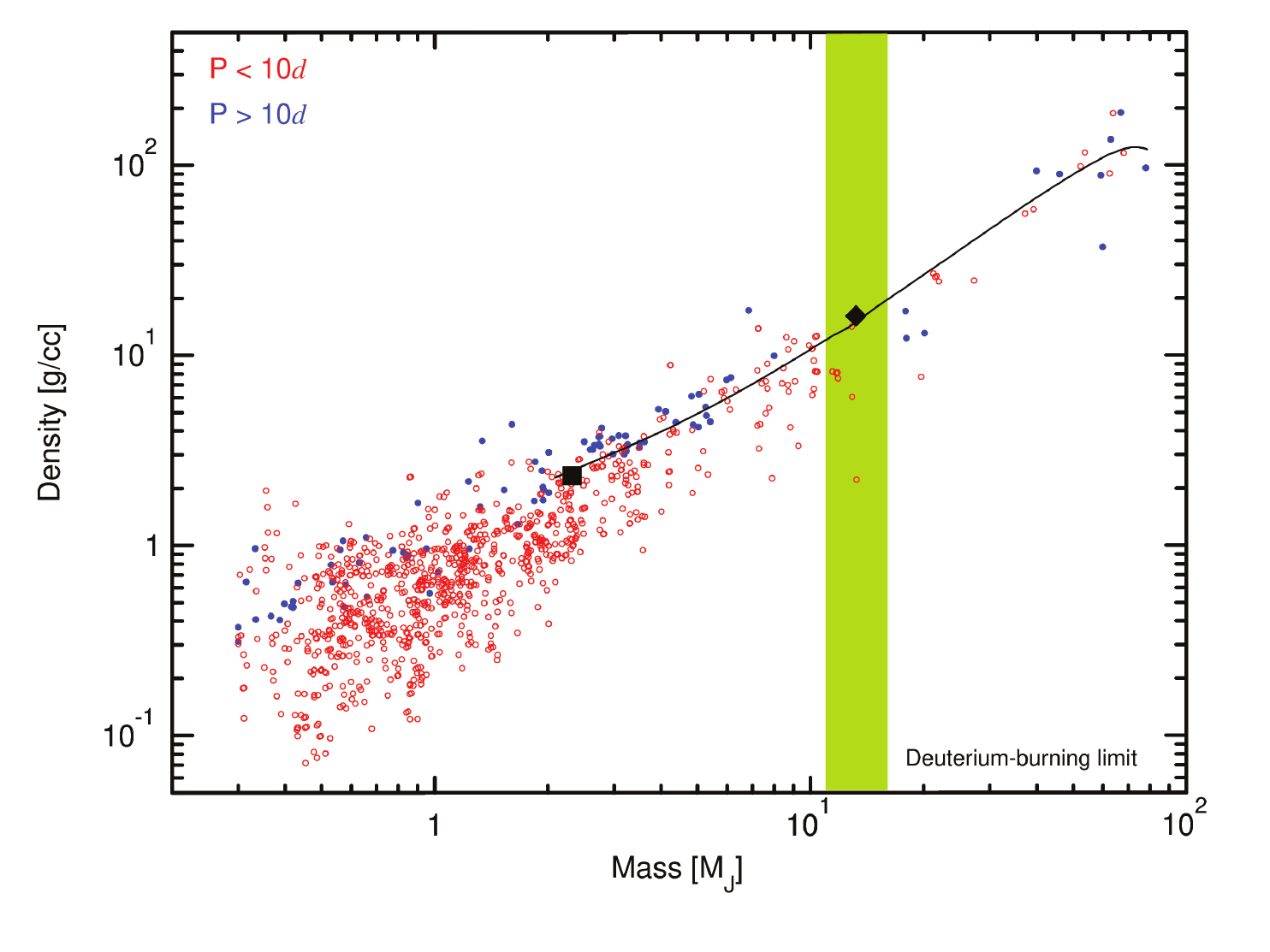}
    \caption{Same as Fig. \ref{fig:mass_rad}, but for the planet density.}
    \label{fig:density_rad}
\end{figure}

\begin{figure}
    \centering
    \includegraphics[angle=90,width=\linewidth]{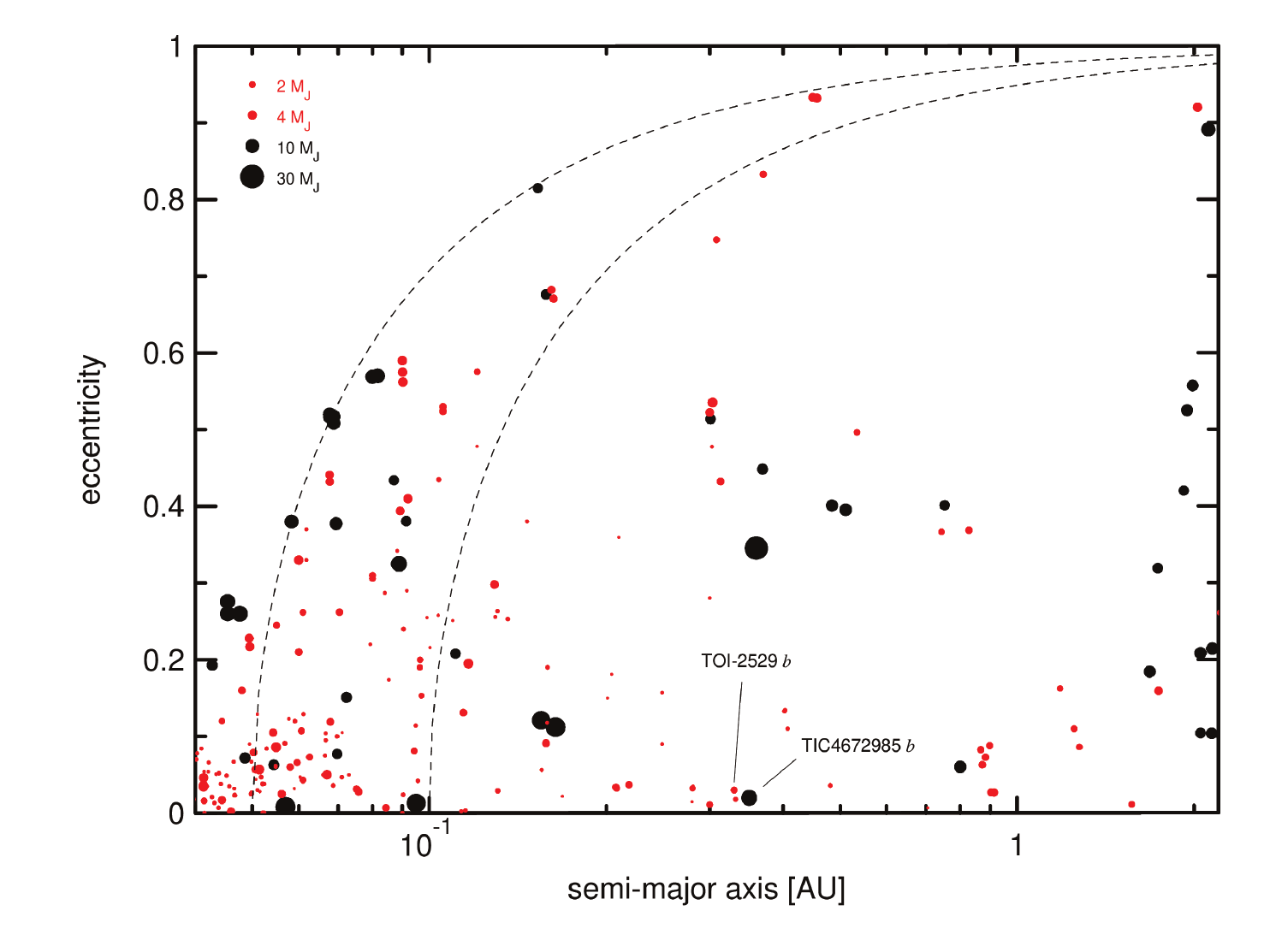}
    \caption{Eccentricity distribution as a function of the orbital distance for known giant planets with eccentricity precision of $\sigma_e <$ 0.15 and mass precision better than 20\%, including TIC\,4672985\,$b$ and TOI-2529\,$b$. The size of the symbols is proportional to the square of the planet's mass. For more clarity, we split the data into two populations, with masses below and above 5 \mjup (red and black dots, respectively). The dotted lines correspond to high-eccentricity migration pathways at constant angular momentum (e.g. \citealt{dong2021}), with final orbital distance of 0.05 and 0.10 AU (left and right curves, respectively).}
    \label{fig:semiaxis_ecc}
\end{figure}

Figure \ref{fig:mass_rad} shows the mass-radius distribution of known\footnote{Data from \href{https://exoplanetarchive.ipac.caltech.edu/}{NASA Exoplanet Archive}, as of June 30, 2023. We complemented this list with known transiting brown-dwarf (BDs) from \citet{Carmichael2021}.}
transiting giant planets, with a mass and radius determination better than 20\%, including  the position of TOI-2529\,$b$ and TIC\,4672985\,$b$. 
For comparison, a solar metallicity, 1 Gyr old ATMO2020 isochrone \citep{phillips2020} is overplotted. The deuterium-burning limit ($\sim$\,11-16 \mjup; \citealt{spiegel2011}) is also shown. TOI-2529\,$b$ is a new example of 
a growing population of well-characterized transiting giant planets with orbital periods $>$ 10 $d$. Because this type of planets has a mild insolation level, they cool down slowly while losing their internal energy in the form of radiation. As a result, and after $\sim$ 1 Gyr from their birth, they shrink to $\sim$ 1 \rjup and reach an effective temperature between $\sim$ 200\,-\,500 K. 
In contrast, closer-in gas giants remain significantly hotter and therefore have larger radii at a given age. These two different populations can be clearly distinguished in Figure \ref{fig:mass_rad}, where hot Jupiters show a radius distribution between $\sim$ 1-2 \rjup, while most warm giant planets pile up around $\sim$ 1 \rjup.
On the other hand, TIC\,4672985\,$b$ is a unique example of a transiting warm planet that is in the transition between the super-Jupiter and the BD regime. Fig. \ref{fig:mass_rad} shows that this object stands alone in the deuterium-burning limit region among the population of warm transiting giant planets. \newline
Similarly, Figure \ref{fig:density_rad} shows the bulk density as a function of 
the planet mass. Again, there is a clear distinction between the hot and warm giant planets. Hot giants present a large scatter, with many examples of extremely low-density objects, as opposed to the warmer counterpart, which mainly follows a linear correlation between these two quantities in log-log space, as predicted by evolutionary models. \newline
Finally, Figure \ref{fig:semiaxis_ecc} shows the eccentricity distribution as a function of the orbital separation. We included only planets with a mass determination\footnote{We included only transiting planets so as to have actual mass measurements rather than RV values of M$_p$\,sin\,$i$ that only provide lower limits to M$_p$} better than 20\% and an eccentricity uncertainty of $\sigma_e <$ 0.15. While TOI\,2529\,$b$ presents a low-eccentricity level, which is common among the population of low-mass giant planets orbiting beyond $\sim$\,0.1\,AU, TIC\,4672985\,$b$ stands out as a unique massive warm substellar companion in a nearly circular orbit.  

\subsection{Internal composition of the planets \label{sec:planet_composition}}

\begin{figure}[ht]
\includegraphics[width=1.20\linewidth,angle=0]{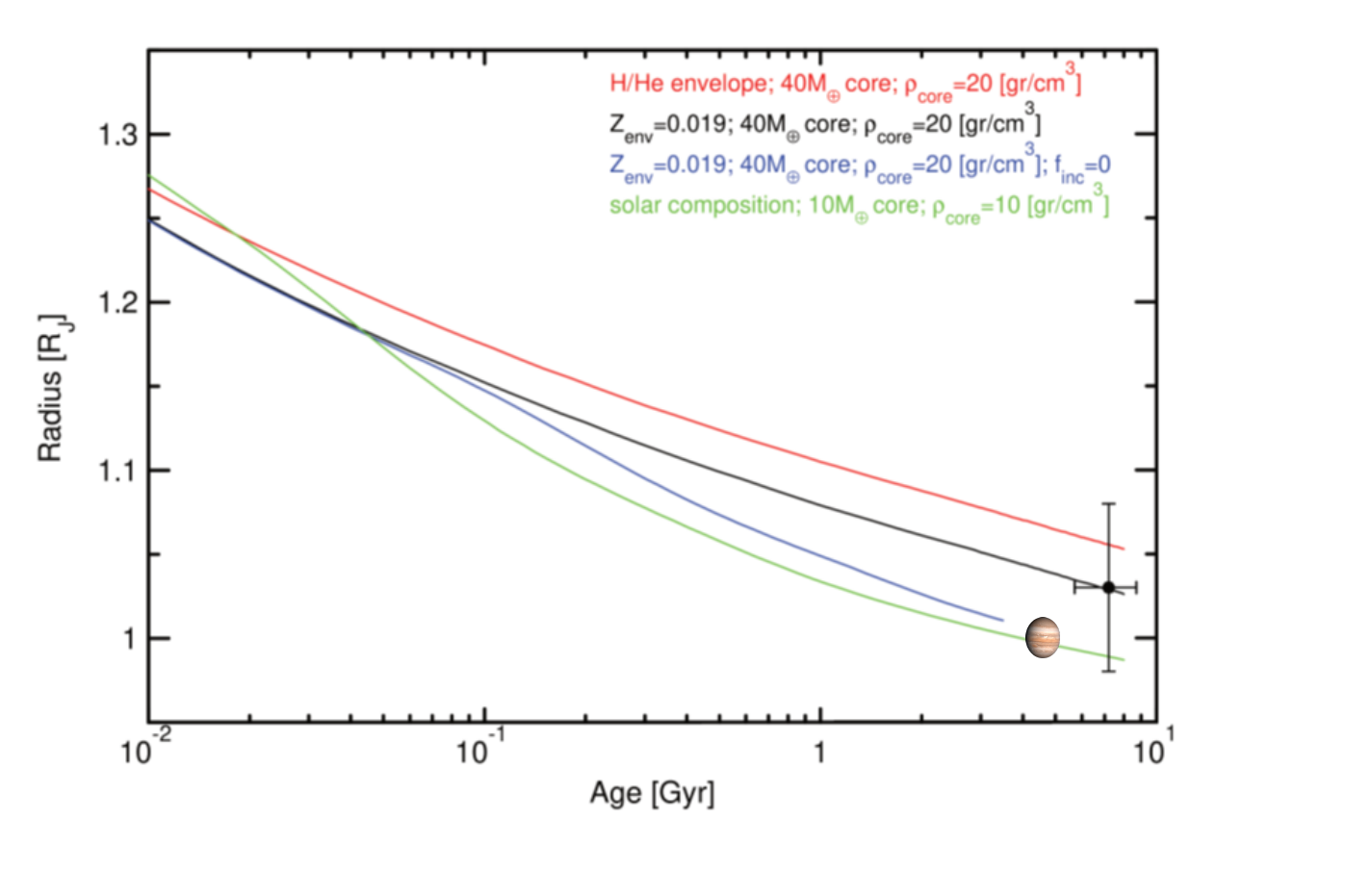}
\caption{Position of TOI-2529\,$b$ in the age-radius diagram (black dot). Planet evolutionary models with different envelope composition and insolation level over-plotted. For comparison, a simple model of Jupiter is also shown.
\label{fig:age_rad_TOI2529}}
\end{figure}

\begin{figure}[ht]
\includegraphics[width=1.18\linewidth,angle=0]{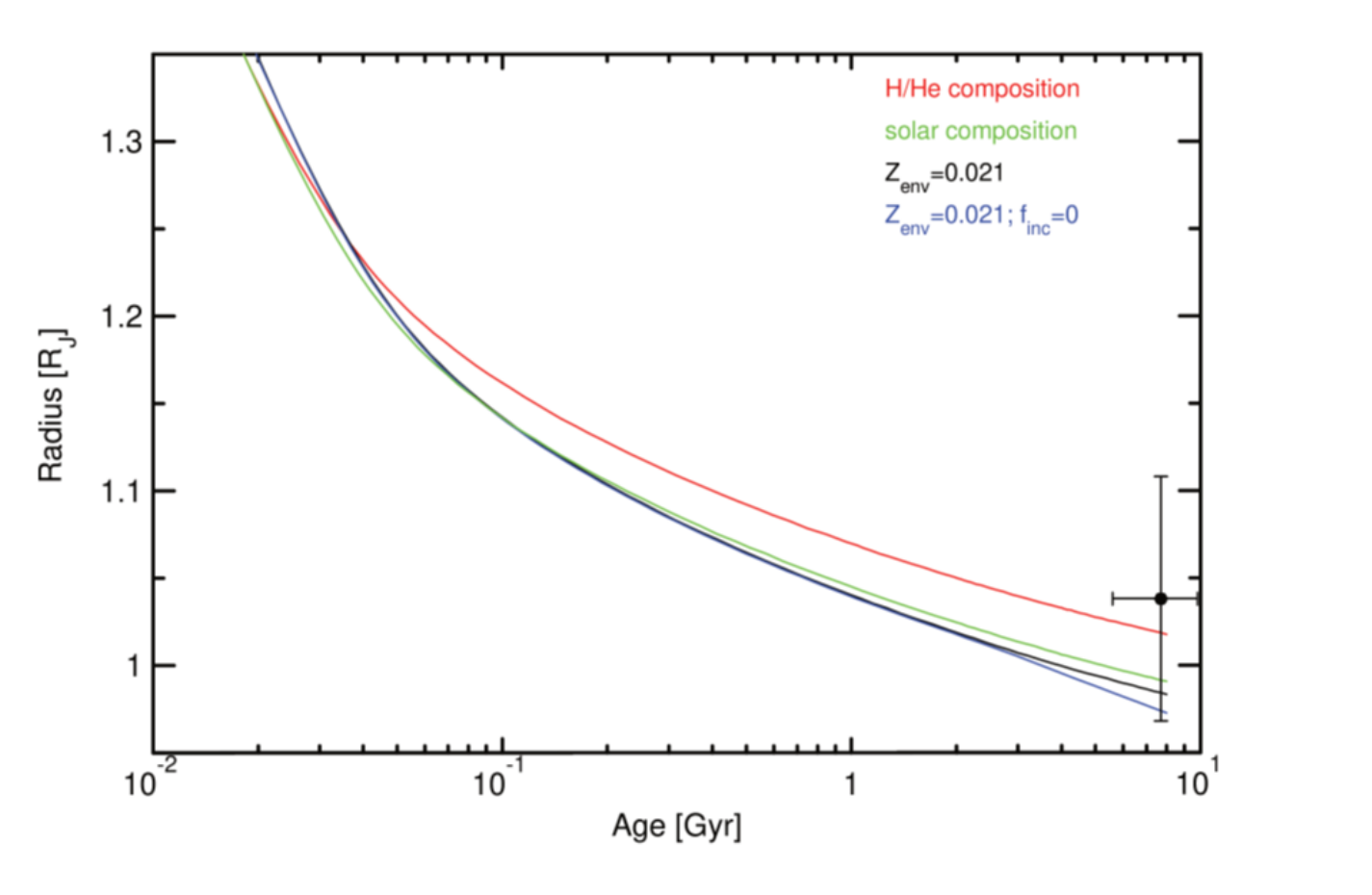}
\caption{
Same as Fig. \ref{fig:age_rad_TOI2529}, but for TIC\,4672985\,$b$. \label{fig:age_rad_TIC4672}}
\end{figure}

To study the composition of the planets presented here, we compared their radii with planet evolutionary models. In particular, we used the modules for experiments in stellar astrophysics (\mesa; \citealt{paxton2011}), and we generated different giant planet models, with a different envelope composition and core mass, and we also included the effect of the stellar irradiation.
To do this, we used the updated\footnote{https://opalopacity.llnl.gov/EOS\_2005} version of the $\rho$-$T$ OPAL tables \citep{OPAL}, with the extension to lower temperatures and densities from the SCvH equation of state (EOS) presented in \citet{saumon1995}. We note that we mimicked the presence of metals in the H/He SCvH EOS by setting an equivalent He content given by Y$^{'}$ = Y + Z, as shown in \citet{chabrier1997}. 
Similarly, we employed the low-temperature opacity tables from \citet{ferguson2005} and \citet{freedman2008}, coupled with the heavy element distribution of \citet{grevesse1998}. These tables also include the radiative opacities from molecules and grains. 
Finally, we adopted the solar metallicity Z = 0.0152 with the corresponding helium-scaled mass fraction: Y = 0.2485 + 1.78 Z, 
derived in \citet{bressan:2012}, which is the one used by \zaspe (see Section \ref{sec:stellar_par}). \newline
We first created an adiabatically contracting model using the {\texttt{create$\_$initial$\_$model}} routine \citep{paxton2013}, which builds a planetary model with a given initial radius and gas mass. 
In our calculations, we set the initial radius to 5 \rjup for all the different models, with the possibility of including an inert rocky core with constant density.
Finally, we evolved the different models for 8 Gyr, and also included the planet insolation. To do this, we implemented the {\texttt{gray\_irradiated}} option inside the {\texttt{atm}} module, which uses the angle-averaged temperature profile described in \citet{guillot2010}. This option includes both the stellar irradiation and the cooling flux from the planetary interior.  
Instead of using the current insolation value, we computed the incident flux received by the planet, by fixing the orbital distance to its current value, and we updated the host star luminosity in steps of 1 Gyr, using \parsec stellar evolutionary models that matched the stellar parameters for each star, as derived here (see Table \ref{Tab:atm_par}). This is particularly important for TOI-2529, which is in the subgiant phase, and therefore, its luminosity has increased significantly since the zero-age main sequence. \newline
To study the composition of TOI-2529\,$b$, we computed different models with a total mass of 2.34 \mjup. In our calculations, we included a coreless and a rocky core model with a different amount of metals in the H/He gas envelope (Z$_{\rm env}$).
Finally, we assumed that the age of the planet is the same as that of the host star, considering the very short lifetime of protoplanetary disks (e.g., \citealt{mamajek2009}). 
Figure \ref{fig:age_rad_TOI2529} shows the position of TOI-2529\,$b$ in the age-radius diagram. Two different models with a 40 \mearth core with constant density of 20 [gr\,cm$^{-3}$] and with different envelope compositions are overplotted. 
For comparison, a model with incident flux (f$_{\rm inc}$) equal to zero is also shown. A possible model that reproduces the current radius of the planet well is the model with a rocky core, Z$_{\rm env}$ = 0.019 (the same as the host star) and with stellar irradiation. 
However, we note that different combinations of the core composition and Z$_{\rm env}$ can reproduce observed properties of the planet within the observational uncertainties in age and radius. 
By fixing the Z$_{\rm env}$ to 0.019 and the density of the core to 20 [gr\,cm$^{-3}$], we found that core masses up to $\sim$\,100\,\mearth
match the observed age-radius of the planet within 1$\sigma$. This corresponds to a heavy element enrichment of $Z_p/Z_\star$\,$\sim$\,1-8.
We also note that even though TOI-2529\,$b$ is a WJ, the effect of the stellar irradiation plays an important role in its cooling evolution, and thus in its radius as a function of time. After $\sim$ 1 Gyr, the effect of the delayed cooling due to the incident flux clearly leads to a radius tthat is $\sim$ 4\% larger than in the model without incident flux (blue line in Figure \ref{fig:age_rad_TOI2529}). 
Moreover, heat deposition from the stellar flux in the planet interior (e.g. \citealt{Ginzburg2016}), would further slow the planet cooling rate down, and thus models with larger heavy element fraction would fit the observed radius of the planet better. Finally, and for comparison, a simple model of Jupiter (age = 4.6 Gyr) is overplotted (green line). \newline
Similarly, Figure \ref{fig:age_rad_TIC4672} shows the position of TIC\,4672985\,$b$ in the age-radius diagram. Again, different models with different compositions, and including the stellar irradiation are shown. For comparison, we also included a model with f$_{\rm inc}$ = 0. 
We note that for this object we did not include a rocky core and we neglected the additional heating from deuterium burning, which can increase the luminosity and halt its contraction at early epochs ($\lesssim$ 50\,-\,100 Myr; e.g. \citealt{Chabrier2000}; \citealt{paxton2011}; \citealt{Molliere2012}), but it has a negligible effect in the long-term evolution of the planet. 
Although a model with the same composition as of the host star matches the current position of TIC\,4672985\,$b$ at the 1$\sigma$ level, models with subsolar metallicity are clearly preferred. This means that no heavy-elements enrichment with respect to the parent star need be invoked.  

\section{Summary and conclusions \label{sec:conclusions}}

We used space- and ground-based 
photometry from different facilities, combined with precision RVs obtained from spectroscopic data collected with different instruments, to characterize two transiting giant planets orbiting the subgiant star TOI-2529 and the main-sequence star TIC\,4672985. From the joint analysis we derived the following parameters for TOI-2529\,$b$: $P$ = \pertwo d, M$_p$ = \masstwo \mjup, R$_p$ = \radtwo \rjup, and $e$ = \ecctwo. 
With an equilibrium temperature of $\sim$ 636 K due to a relatively low current insolation of 3.7$\times$10$^7$ [erg\,s$^{-1}$\,cm$^{-2}$], this planet is an excellent example of an emerging population of very well characterized transiting WJs. 
In terms of the orbital configuration, TOI-2529\,$b$ is found in a nearly circular orbit, which is consistent with an in situ formation channel via core-accretion (e.g., \citealt{Bodenheimer2000}; \citealt{Boley2016}; \citealt{Batygin2016}) or by a type II disk migration pathway (e.g. \citealt{Lin1993}). In this scenario, even in the case of a Jupiter-mass planet in a highly inclined and eccentric orbit (e.g., due to planet-planet interactions; \citealt{Weidenschilling1996}), the tidal interactions with the disk lead to a rapid damping (on timescales of about 10$^3$ yr) of the inclination and the eccentricity \citep{Marzari2009}.
In addition, TOI-2529\,$b$ orbits a metal-rich star, following the well-known planet-metallicity correlation for gas giants orbiting FGK dwarf stars (\citealt{Gonzalez1997}; \citealt{Santos2004}; \citealt{Fischer2005}; \citealt{Adibekyan2019}; \citealt{Osborn2020}), which is also valid for more massive evolved stars (e.g., \citealt{Jones2016}). This correlation provides strong observational support for the core-accretion formation model.
Furthermore, based on our calculations, the planet composition is consistent with a rocky core and heavy element enrichment with respect to its parent star, providing further evidence of a formation channel via core accretion.
Given its low equilibrium temperature and relatively compact radius, TOI-2529\,$b$ has a very low transmission spectroscopy metric (TSM; \citealt{Kempton2018}) of $\sim$ 5, making this a very challenging system for a detailed atmospheric characterization study. 
On the other hand, we estimated a Rossiter-McLaughlin (R-M; \citealt{Rossiter1924}; \citealt{McLaughlin1924}) RV amplitude of $\sim$ 7 \ms (assuming $i_\star$ = 90 deg; see eq. (1) in \citealt{Triaud2018}), which is within the reach of stable instruments mounted on 8-meter-class telescopes (due to the relatively faintness of the target), such as ESPRESSO \citep{Pepe2021}. A low obliquity level would provide further observational support for the disk migration scenario (e.g \citealt{Lin1996}), highlighting the importance of a future R-M characterization of this system. \newline
For TIC\,4672985\,$b$ we obtained the following orbital parameters: $P$ = \perone d, M$_p$ = \massone \mjup, R$_p$ = \radone \rjup, and $e$ = \eccone.
Additionally, we detected an RV trend at the $\sim$\,350 \ms yr$^{-1}$ level, indicating the presence of either an outer BD companion in the system or a bound low-luminosity stellar companion.
Thus, this is a unique system. It is the first warm transiting substellar companion that is located in the regime of super-Jupiters to brown dwarfs and has a precise mass and radius determination. 
Interestingly, TIC\,4672985\,$b$ also presents a very low eccentricity, which is extremely rare among planets of its class. Moreover, this observational result challenges theoretical migration models, where unlike the case of lower-mass gas giants, a low eccentricity like this is not a direct indication of coplanar migration via disk interactions because even at low inclinations with respect to the disk, the eccentricity can be excited by this mechanism in this planetary mass regime (\citealt{Bitsch2013}; \citealt{Dunhill2013}). 
Similarly, the presence of an outer companion might induce eccentricity growth via Kozai-Lidov cycles (\citealt{kozai1962}; \citealt{lidov1962}) or through nearly coplanar apsidal precession resonances (e.g., \citealt{Anderson2017}). 
Measuring the obliquity level for this object (predicted R-M amplitude of $\sim$ 8 \ms) is thus mandatory to further characterize the orbital configuration of the system. 
On the other hand, and due to its very low TSM of $\sim$ 1, it is not possible to perform a detailed atmospheric study of this planet. \newline
Finally, from our calculations, the position of TIC\,4672985\,$b$ in the age-radius diagram is consistent with a coreless model with subsolar metallicity, meaning that no metal enhancement with respect to the parent star is present. This could be an indication that this object was formed by gravitational instability (e.g., \citealt{Boss1997}; \citealt{Bate2012}), as was shown to be the case for substellar objects more massive than $\sim$ 10\,\mjup \citep{schlaufman2018}.

\begin{acknowledgements}

The results reported herein benefited from collaborations and/or information exchange within NASA’s Nexus for Exoplanet System Science (NExSS) research coordination network sponsored by NASA’s Science Mission Directorate under Agreement No. 80NSSC21K0593 for the program “Alien Earths”.
We thank the Swiss National Science Foundation (SNSF) and the Geneva University for their continuous support to the planet search programs. 
This work has been in particular carried out in the frame of the National Centre for Competence in Research PlanetS supported by the SNSF under grants 51NF40\_182901 and 51NF40\_205606
This publication made use of The Data \& Analysis Center for Exoplanets (DACE), which is a facility based at the University of Geneva dedicated to extrasolar planets data visualisation, exchange and analysis. DACE is a platform of NCCR $PlanetS$ and is available at https:$//$dace.unige.ch. 
This paper made use of data collected by the TESS mission and are publicly available from the Mikulski Archive for Space Telescopes (MAST) operated by the Space Telescope Science Institute (STScI). Funding for the TESS mission is provided by NASA’s Science Mission Directorate. 
We acknowledge the use of public TESS data from pipelines at the TESS Science Office and at the TESS Science Processing Operations Center. Resources supporting this work were provided by the NASA High-End Computing (HEC) Program through the NASA Advanced Supercomputing (NAS) Division at Ames Research Center for the production of the SPOC data products.
This paper is based on data collected by the SPECULOOS-South Observatory at the ESO Paranal Observatory in Chile. The ULiege's contribution to SPECULOOS has received funding from the European Research Council under the European Union's Seventh Framework Programme (FP/2007-2013) (grant Agreement n$^\circ$ 336480/SPECULOOS), from the Balzan Prize and Francqui Foundations, from the Belgian Scientific Research Foundation (F.R.S.-FNRS; grant n$^\circ$ T.0109.20), from the University of Liege, and from the ARC grant for Concerted Research Actions financed by the Wallonia-Brussels Federation. 
The Cambridge contribution is supported by a grant from the Simons Foundation (PI Queloz, grant number 327127). The Birmingham contribution to SPECULOOS is in part funded by the European Union's Horizon 2020 research and innovation programme (grant's agreement n$^{\circ}$ 803193/BEBOP), from the MERAC foundation, and from the Science and Technology Facilities Council (STFC; grant n$^\circ$ ST/S00193X/1, and ST/W000385/1).
This publication benefits from the support of the French Community of Belgium in the context of the FRIA Doctoral Grant awarded to MT. ML acknowledges support of the Swiss National Science Foundation under grant number PCEFP2\_194576.
MRS acknowledges support from the UK Science and Technology Facilities Council (ST/T000295/1) and support from the European Space Agency as an ESA Research Fellow.
MTP acknowledges the support of the Fondecyt-ANID Post-doctoral fellowship No. 3210253.
TT acknowledges support by the DFG Research Unit FOR 2544 "Blue Planets around Red Stars" project No. KU 3625/2-1.
TT further acknowledges support by the BNSF program "VIHREN-2021" project No. KP-06-DV/5. 
DD acknowledges support from the NASA Exoplanet Research Program grant 18-2XRP18\_2-0136, and from the TESS Guest Investigator Program grants 80NSSC22K0185 and 80NSSC23K0769. JC acknowledges support of the South Carolina Space Grant Consortium. RB acknowledges support from FONDECYT project 11200751 and from project IC120009 “Millennium Institute of Astrophysics (MAS)” of the Millenium Science Initiative. MU gratefully acknowledges funding from the Research Foundation Flanders (FWO) by means of a junior postdoctoral fellowship (grant agreement No. 1247624N).
\end{acknowledgements}

\bibliographystyle{aa}
\bibliography{Two_massive_warm_Jupiters}{}

\begin{appendix} 
\onecolumn
\section{Radial velocity tables.}

\begin{longtable}{l c c c c r}
\caption{Relative radial velocities for TIC\,4672985. \label{tic4672_rv} }\\
\hline \hline 
\vspace{-0.3cm} \\
BJD-2450000 & RV (\ms) & $\sigma_{RV}$ (\ms) & BVS (\ms) & $\sigma_{BVS}$ (\ms) & Instrument \\
\hline \vspace{-0.3cm}  \\
\endhead
\hline
\multicolumn{6}{r}{\footnotesize\itshape Continue on the next page}
\endfoot
\hline
\endlastfoot
  59222.54345 &     162.4 &    10.7 &   116.8 &    47.8 & CHIRON \\
  59225.59389 &     317.1 &    11.1 &    69.5 &    39.6 & CHIRON \\
  59229.60416 &     543.4 &    13.6 &    37.8 &    40.9 & CHIRON \\
  59234.56679 &     736.5 &    14.8 &    -2.3 &    44.5 & CHIRON \\
  59400.91392 &    -229.2 &    15.0 &   -65.8 &    54.0 & CHIRON \\
  59400.93624 &    -230.3 &    10.1 &    -1.0 &    38.5 & CHIRON \\
  59412.91660 &    -589.7 &    12.7 &     8.4 &    43.6 & CHIRON \\
  59417.83555 &    -567.8 &    13.0 &    -5.8 &    33.2 & CHIRON \\
  59422.89921 &    -364.9 &    10.9 &    16.6 &    39.8 & CHIRON \\
  59448.85994 &     619.4 &     9.7 &   -40.2 &    32.1 & CHIRON \\
  59452.81067 &     576.3 &    15.1 &    36.1 &    59.5 & CHIRON \\
  59459.83254 &     308.3 &    12.8 &  -156.0 &    61.5 & CHIRON \\
  59465.84358 &     -71.6 &    29.3 &   -79.6 &   122.5 & CHIRON \\
  59480.68472 &    -628.4 &    20.4 &    43.9 &    56.3 & CHIRON \\
  59489.73775 &    -467.5 &    25.4 &    65.5 &    73.7 & CHIRON \\
  59498.71994 &     -14.4 &    23.3 &   123.9 &    62.0 & CHIRON \\
  59501.70927 &     125.8 &    26.0 &    97.5 &    78.5 & CHIRON \\
  59504.66754 &     277.7 &    15.8 &   -59.1 &    39.9 & CHIRON \\
  59526.58391 &     354.7 &     9.3 &    50.0 &    37.3 & CHIRON \\
  59531.66372 &      70.9 &    12.1 &    17.0 &    28.2 & CHIRON \\
  59536.60366 &    -233.1 &    10.4 &     9.9 &    33.8 & CHIRON \\
  59593.53740 &     396.2 &     9.6 &    69.2 &    56.6 & CHIRON \\
  59846.76957 &    -264.5 &    10.5 &     3.6 &    41.2 & CHIRON \\
  59859.69126 &     264.4 &    12.4 &   -29.4 &    37.4 & CHIRON \\
  60179.88229 &   -1090.9 &    19.4 &     2.0 &    64.7 & CHIRON \\
 59188.55229  & 30419.6 &   7.2 & -20.0 &  11.0 & FEROS \\
 59191.58077  & 30238.5 &   6.8 &   0.0 &  10.0 & FEROS \\
 59194.58783  & 30068.3 &   7.3 &  -8.0 &  11.0 & FEROS \\
 59196.54313  & 29967.6 &   6.1 &  -8.0 &   9.0 & FEROS \\
 59206.55663  & 29736.2 &   8.8 & -15.0 &  12.0 & FEROS \\
 59212.57083  & 29818.2 &   7.5 &  -3.0 &  11.0 & FEROS \\
 59216.65293  & 29987.7 &   7.9 &  -6.0 &  11.0 & FEROS \\
 59218.57604  & 30070.2 &   6.6 &  -6.0 &  10.0 & FEROS \\
 59222.61173  & 30300.8 &   6.9 &  -4.0 &  10.0 & FEROS \\
 59223.63904  & 30323.7 &   8.0 & -38.0 &  11.0 & FEROS \\
 59422.85717  & 29733.6 &   7.0 &  10.0 &  10.0 & FEROS \\
 59482.72687  & 29472.6 &   6.5 &  -6.0 &  10.0 & FEROS \\
 59484.79610  & 29479.8 &   6.4 &   0.0 &  10.0 & FEROS \\
 59486.77790  & 29536.5 &   6.5 & -16.0 &  10.0 & FEROS \\
 59497.85912  & 30013.9 &   7.5 & -12.0 &  11.0 & FEROS \\
 59502.77381  & 30248.3 &   7.1 &  -2.0 &  11.0 & FEROS \\
 59540.63774  & 29681.6 &   7.0 &  22.0 &  10.0 & FEROS \\
 59546.55133  & 29490.0 &   8.5 & -28.0 &  12.0 & FEROS \\
 59553.59297  & 29417.6 &   7.3 & -11.0 &  11.0 & FEROS \\
 59587.61015  & 30598.7 &   7.3 & -21.0 &  11.0 & FEROS \\
 59171.68146 & 31052.9 & 12.2 & -25.3 & 15.9 & HARPS \\ 
 59189.61998 & 30366.6 & 12.2 & 5.0   & 15.9 & HARPS \\ 
 59216.61051 & 30014.4 & 6.3  & 22.1  & 8.2  & HARPS \\ 
 59219.54050 & 30136.1 & 8.5  & 4.7   & 11.1 & HARPS \\ 
 59220.55062 & 30202.1 & 7.3  & 10.5  & 9.5  & HARPS \\ 
 59230.54004 & 30708.9 & 6.7  & -18.4 & 8.8  & HARPS \\ 
 59245.52983 & 30917.9 & 10.1 & -21.7 & 13.2 & HARPS \\ 
 59245.54129 & 30930.6 & 5.9  & -23.6 & 7.6  & HARPS \\ 
 59250.53267 & 30747.0 & 4.5  & -2.7  & 5.8  & HARPS \\ 
 59251.53713 & 30718.8 & 7.3  & -11.2 & 9.5  & HARPS \\ 
 59254.52904 & 30533.4 & 5.5  & -4.4  & 7.1  & HARPS \\ 
 59258.54084 & 30318.6 & 6.3  & 1.7   & 8.2  & HARPS \\ 
 59470.82587 & 29821.6 & 7.8  & 34.2  & 10.2 & HARPS \\ 
 59490.73357 & 29642.7 & 9.2  & -16.6 & 12.0 & HARPS \\ 
 59502.79276 & 30257.9 & 7.8  & 3.6   & 10.2 & HARPS \\ 
 59514.73359 & 30692.0 & 5.5  & -1.8  & 7.1  & HARPS \\ 
 59531.60765 & 30218.5 & 7.8  & -7.4  & 10.2 & HARPS \\ 
 59532.66751 & 30166.7 & 7.2  & 15.0  & 9.4  & HARPS \\ 
  59182.63980 &   30755.0 &    22.5 &   -77.1 &    31.9 & CORALIE \\
  59187.68617 &   30491.1 &    30.7 &   -57.1 &    43.4 & CORALIE \\
  59196.58888 &   29998.8 &    22.6 &    -6.7 &    31.9 & CORALIE \\
  59201.63030 &   29789.8 &    33.7 &   -56.6 &    47.7 & CORALIE \\
  59208.59669 &   29750.5 &    34.8 &   -61.9 &    49.2 & CORALIE \\
  59214.57535 &   29920.0 &    28.8 &   -42.3 &    40.7 & CORALIE \\
  59231.56773 &   30737.8 &    47.0 &   -22.3 &    66.4 & CORALIE \\
  59244.56099 &   30930.6 &    42.4 &   -39.8 &    59.9 & CORALIE \\
  59250.56540 &   30696.4 &    45.1 &   -15.0 &    63.8 & CORALIE \\
  59253.53569 &   30561.6 &    41.8 &   -39.7 &    59.1 & CORALIE \\
  59260.54649 &   30203.5 &    49.4 &   -63.1 &    69.9 & CORALIE \\
  59455.76669 &   30582.3 &    24.5 &   -25.7 &    34.7 & CORALIE \\
  59464.80650 &   30117.6 &    26.8 &   -37.1 &    37.9 & CORALIE \\
  59473.70146 &   29639.4 &    27.1 &   -72.8 &    38.3 & CORALIE \\
  59479.73983 &   29479.2 &    27.9 &   -32.4 &    39.5 & CORALIE \\
  59490.66775 &   29650.8 &    30.1 &    15.5 &    42.5 & CORALIE \\
  59504.69109 &   30325.1 &    25.2 &   -65.6 &    35.7 & CORALIE \\
  59512.75189 &   30639.3 &    23.7 &   -35.9 &    33.5 & CORALIE \\
  59521.56897 &   30607.6 &    44.0 &   -45.8 &    62.2 & CORALIE \\
  59542.62956 &   29577.4 &    29.2 &   -53.2 &    41.4 & CORALIE \\
  59580.60906 &   30522.8 &    22.7 &   -73.5 &    32.2 & CORALIE \\
  59609.57162 &   29634.5 &    20.4 &   -17.2 &    28.8 & CORALIE \\
  59825.83112 &   29165.8 &    22.7 &   -45.1 &    32.1 & CORALIE \\
  59826.78639 &   29191.5 &    17.0 &    -6.1 &    24.0 & CORALIE \\
  59846.81771 &   29855.9 &    50.8 &    14.0 &    71.8 & CORALIE \\
  59862.77739 &   30381.1 &    19.1 &   -70.7 &    27.0 & CORALIE \\
  59873.68405 &   30067.9 &    14.7 &     8.5 &    20.8 & CORALIE \\
  59893.66991 &   29137.4 &    18.6 &   -84.8 &    26.3 & CORALIE \\
  59918.58953 &   29950.6 &    12.6 &   -31.7 &    17.8 & CORALIE \\
\hline
\end{longtable}

\begin{table}
\centering
\caption{Relative radial velocities for TOI-2529. \label{tic2693_rv}}
\begin{tabular}{lrrrrr}
\hline\hline
\vspace{-0.3cm} \\
BJD-2450000 & RV (\ms) & err (\ms) & BVS (\ms) & eBVS (\ms) & instrument\\
\hline \vspace{-0.3cm}    
\\
 59239.70586 &   46.5 & 10.3 &   33.1 &  44.2 & CHIRON \\
 59247.70453 &   -5.1 & 10.1 &   90.3 &  32.6 & CHIRON \\
 59249.60260 &  -24.7 & 12.8 &   41.4 &  35.8 & CHIRON \\
 59255.56122 & -123.4 & 11.3 &   -0.8 &  32.3 & CHIRON \\
 59269.67164 & -132.0 &  8.6 &   31.8 &  27.6 & CHIRON \\
 59279.55943 &  -79.3 & 12.4 &  -49.1 &  17.7 & CHIRON \\
 59291.66346 &   30.0 & 12.3 &  -37.1 &  28.5 & CHIRON \\
 59305.56255 &    6.8 & 15.2 &  -42.5 &  31.9 & CHIRON \\
 59345.50162 &  -18.4 & 13.3 &   16.8 &  38.8 & CHIRON \\
 59360.47119 &   17.0 & 12.7 &  -73.6 &  29.0 & CHIRON \\
 59474.86950 &   10.1 &  9.4 &   84.3 &  32.2 & CHIRON \\
 59483.88722 &   94.2 & 18.7 &   11.1 &  38.4 & CHIRON \\
 59491.85983 &  134.2 & 21.7 &  149.3 &  88.6 & CHIRON \\
 59498.87897 &   47.8 & 34.3 &  -29.0 &  74.9 & CHIRON \\
 59536.80881 &  -43.7 &  9.3 &  -12.1 &  22.9 & CHIRON \\
 59544.82069 &   -4.8 &  7.8 &   51.1 &  21.2 & CHIRON \\
 59551.73266 &   55.9 & 10.6 &   31.7 &  27.9 & CHIRON \\
 59559.83665 &   71.6 & 12.4 &   53.9 &  36.3 & CHIRON \\
 59566.73667 &    6.2 &  9.3 &   35.3 &  29.3 & CHIRON \\
 59597.72786 &  -65.5 & 13.2 &   -4.2 &  37.6 & CHIRON \\
 59603.60796 &  -24.0 &  7.6 &   14.0 &  20.8 & CHIRON \\
 59187.75268 & 21240.8 & 7.9 & 0.0 & 11.0 & FEROS \\
 59191.80547 & 21221.9 & 7.5 & 9.0 & 11.0 & FEROS \\
 59193.74595 & 21201.2 & 7.8 & 2.0 & 11.0 & FEROS \\
 59195.72076 & 21202.5 & 8.4 & 8.0 & 11.0 & FEROS \\
 59205.71383 & 21219.1 & 8.3 & -6.0 & 11.0 & FEROS \\
 59209.73037 & 21221.3 & 8.9 & 9.0 & 12.0 & FEROS \\
 59216.75761 & 21286.7 & 8.6 & -4.0 & 12.0 & FEROS \\
 59218.74168 & 21292.8 & 7.8 & 18.0 & 11.0 & FEROS \\
 59220.80439 & 21323.5 & 10.5 & 3.0 & 14.0 & FEROS \\
 59223.82981 & 21359.8 & 8.7 & 7.0 & 12.0 & FEROS \\
 59265.68708 & 21148.6 & 7.7 & 12.0 & 11.0 & FEROS \\
 59273.72703 & 21219.5 & 9.6 & 40.0 & 13.0 & FEROS \\
 59277.53902 & 21222.1 & 8.2 & 11.0 & 11.0 & FEROS \\
 59281.70796 & 21246.1 & 8.8 & -28.0 & 12.0 & FEROS \\
 59283.71154 & 21280.7 & 8.2 & 5.0 & 11.0 & FEROS \\
 59577.78443 & 21227.7 & 9.4 & 1.0 & 12.0 & FEROS \\
 59578.78822 & 21206.3 & 7.8 & 11.0 & 11.0 & FEROS \\
 59227.25224 & 21389.9 & 3.8 & 22.0 & 5.0 & HARPS \\
 59231.23038 & 21406.4 & 3.6 & 18.0 & 5.0 & HARPS \\
 59234.21435 & 21402.2 & 3.8 & 12.0 & 5.0 & HARPS \\
 59238.21866 & 21382.9 & 5.6 & 10.0 & 7.0 & HARPS \\
 59244.20042 & 21344.0 & 3.8 & 36.0 & 5.0 & HARPS \\
 59247.19025 & 21321.2 & 4.9 & 14.0 & 6.0 & HARPS \\
 59254.20689 & 21206.1 & 3.6 & 26.0 & 5.0 & HARPS \\
 59256.16169 & 21224.0 & 4.0 & 15.0 & 5.0 & HARPS \\
 59260.15700 & 21189.4 & 4.3 & 6.0 & 6.0 & HARPS \\
 59263.19458 & 21175.3 & 5.2 & 19.0 & 7.0 & HARPS \\
 59268.17774 & 21194.2 & 6.4 & 16.0 & 8.0 & HARPS \\
 59280.11283 & 21273.6 & 5.2 & 37.0 & 7.0 & HARPS \\
 59291.16541 & 21372.6 & 4.9 & 16.0 & 6.0 & HARPS \\
 59295.17803 & 21403.2 & 11.2 & 39.0 & 15.0 & HARPS \\
 59364.98782 & 21413.3 & 5.2 & 6.0 & 7.0 & HARPS \\
 59369.00836 & 21371.7 & 7.4 & 4.0 & 10.0 & HARPS \\
 59371.00808 & 21369.3 & 8.6 & 26.0 & 11.0 & HARPS \\
 59372.94187 & 21374.3 & 8.0 & 3.0 & 10.0 & HARPS \\
 59505.36859 & 21341.4 & 5.9 & 1.0 & 8.0 & HARPS \\
 59506.36871 & 21300.0 & 5.2 & 22.0 & 7.0 & HARPS \\
 59518.32891 & 21206.2 & 4.8 & 16.0 & 6.0 & HARPS \\
 59533.28374 & 21230.7 & 4.9 & 21.0 & 6.0 & HARPS \\
 59562.26679 & 21365.7 & 10.3 & -4.0 & 13.0 & HARPS \\
 59577.24623 & 21238.3 & 3.8 & 23.0 & 5.0 & HARPS \\
 59622.21689 & 21405.1 & 4.9 & 23.0 & 6.0 & HARPS \\
 59639.21249 & 21269.2 & 6.0 & 8.0 & 8.0 & HARPS \\
 59658.19716 & 21191.5 & 8.0 & -2.0 & 10.0 & HARPS \\
\vspace{-0.3cm} \\
\hline\hline
\end{tabular}
\end{table}

\newpage

\begin{figure*}[!h]
       \includegraphics[angle=90,width = 0.7\columnwidth]{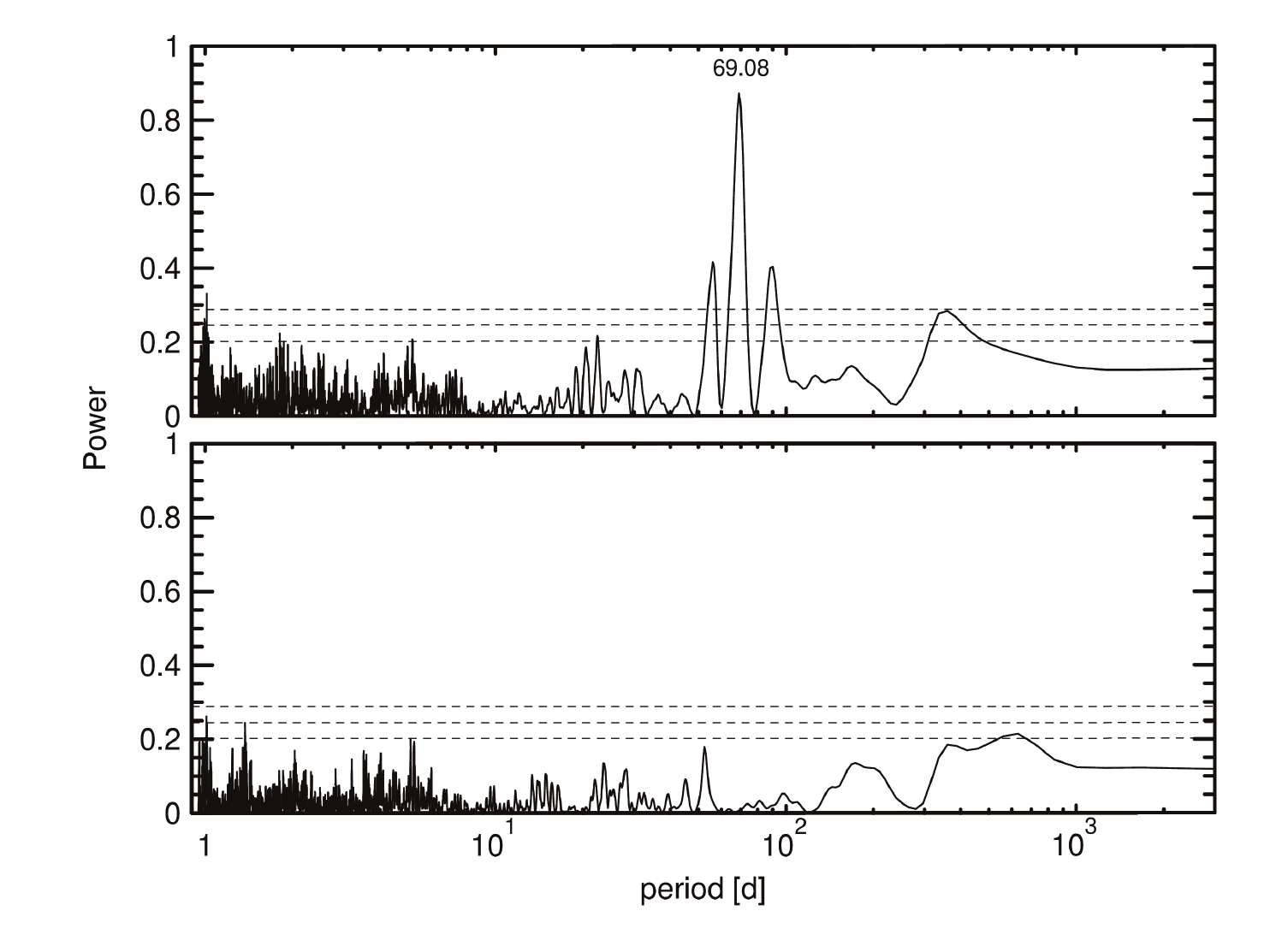}
    \caption{Upper panel: Periodogram of the combined RVs of TIC\,4672985, after correcting by the long-term quadratic trend and applying the differential instrumental zero points. The dashed horizontal lines correspond to 10\%, 1\% and 0.1\% false-alarm probability (from bottom to top). The period of maximum power is labeled. Lower panel: Periodogram of the post-fit residuals.    \label{tic4627_periodogram}}
\end{figure*}

\begin{figure*}[!h]
       \includegraphics[angle=90,width = 0.7\columnwidth]{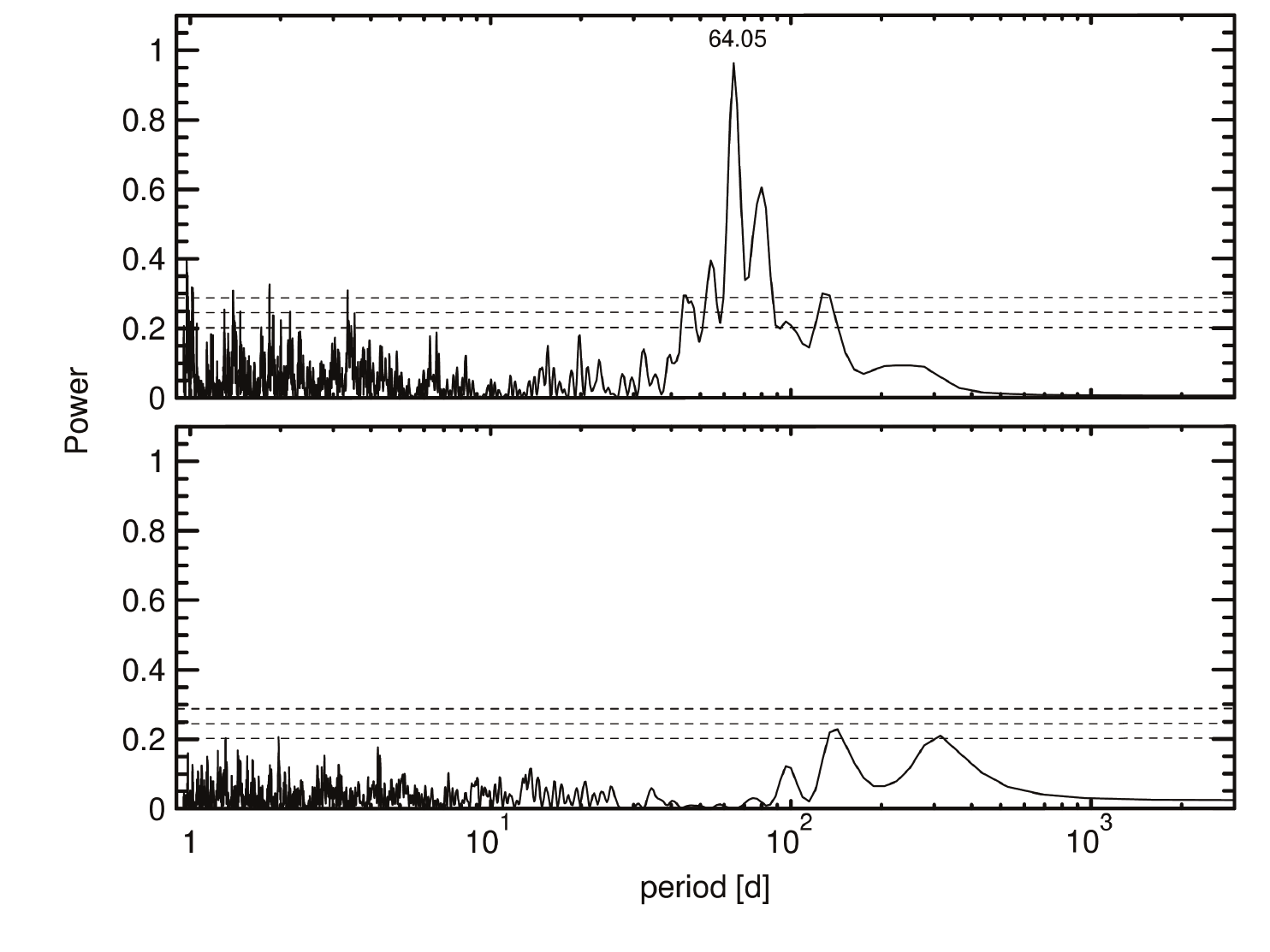}
    \caption{Same as for Fig. \ref{tic4627_periodogram}, but for TOI-2529.  \label{toi2529_periodogram}}
\end{figure*}

\begin{figure*}[!h]
       \includegraphics[angle=90,width = 0.7\columnwidth]{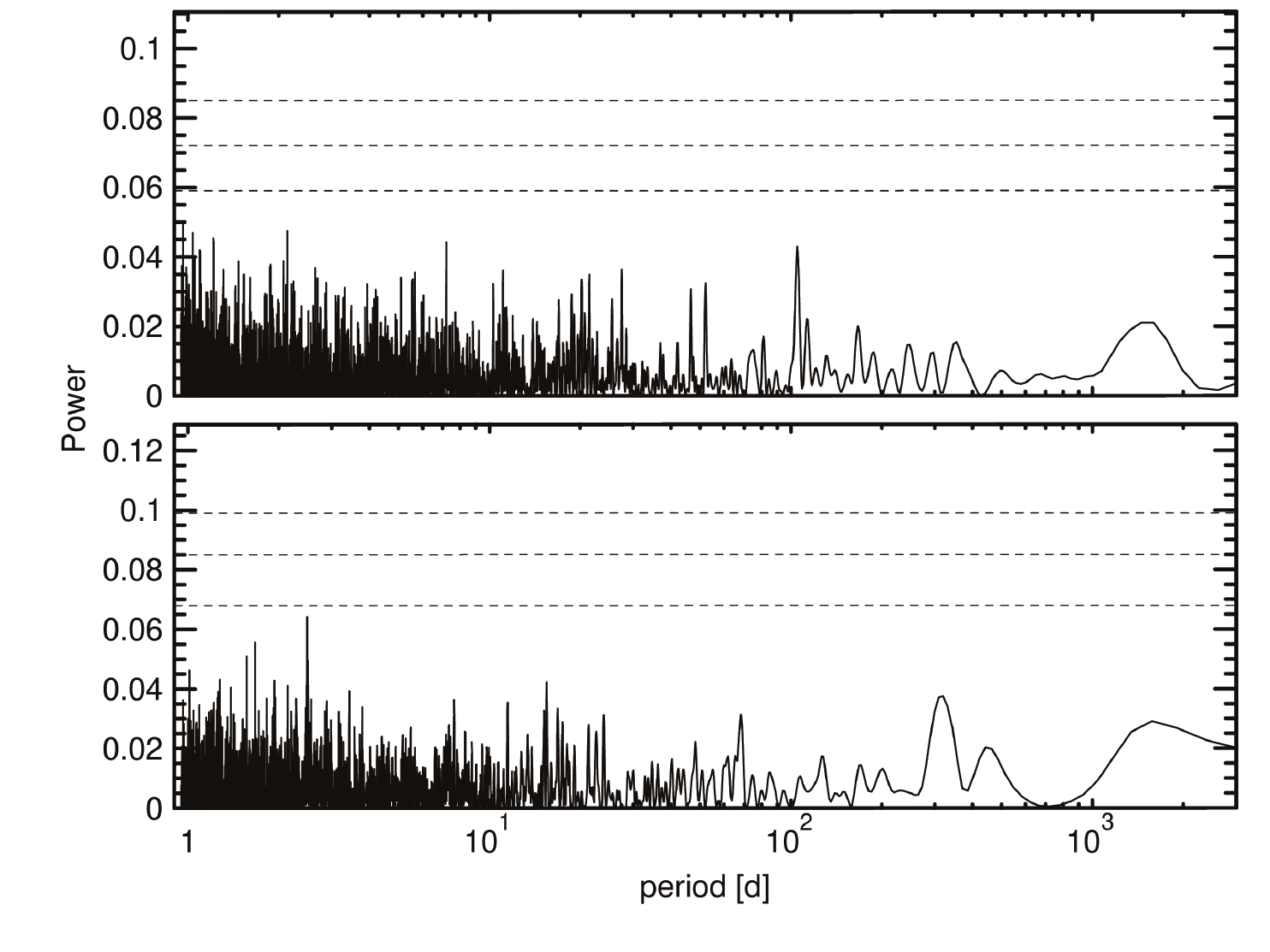}
    \caption{Periodogram of the ASAS and ASAS-SN V-band photometry of TIC\,4672985 (upper and lower panel, respectively). The dashed horizontal lines correspond to 10\%, 1\% and 0.1\% false-alarm probability (from bottom to top). \label{tic4672_photometry_periodogram}}
\end{figure*}

\begin{figure*}[!h]
       \includegraphics[angle=90,width = 0.7\columnwidth]{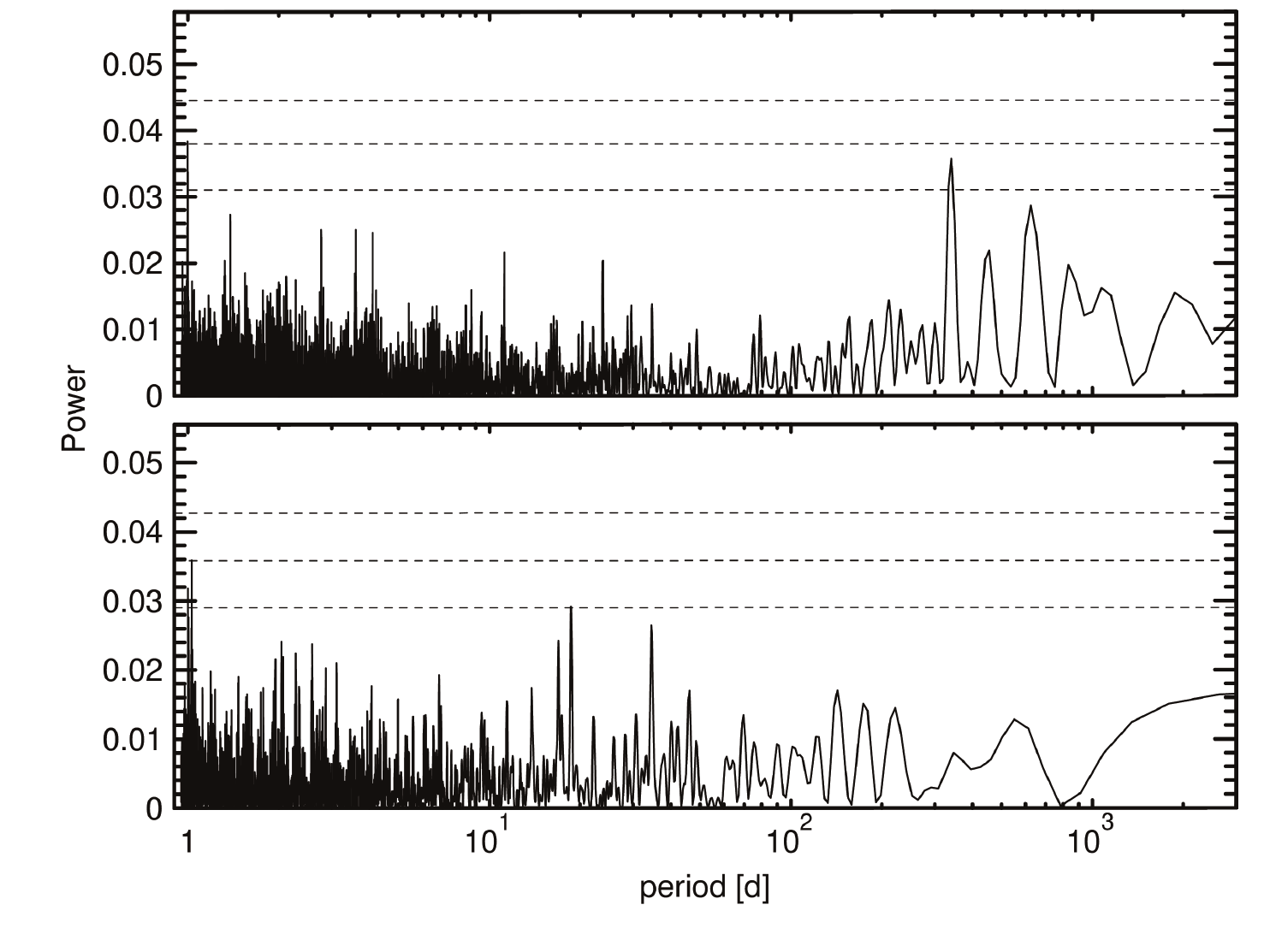}
    \caption{Same as for Fig. \ref{tic4672_photometry_periodogram}, but for TOI-2529.     \label{toi2529_photometry_periodogram}}
\end{figure*}





\end{appendix}

\end{document}